\documentclass[%
twocolumn, prb, aps, superscriptaddress, longbibliography,showpacs,amsmath,amssymb,floatfix
]{revtex4-2}

\usepackage[colorlinks,
            linkcolor=blue,
            anchorcolor=blue,
            citecolor=blue]{hyperref}
\usepackage{graphicx}
\usepackage{dcolumn}
\usepackage{xcolor}
\usepackage{multirow}
\usepackage{bm}

\begin{document}

\preprint{APS/123-QED}

\title{The general approach to the critical phase with coupled quasiperiodic chains}

\author{Xiaoshui Lin}
\affiliation{CAS Key Laboratory of Quantum Information, University of Science and Technology of China, Hefei, 230026, China}
\author{Xiaoman Chen}
\affiliation{CAS Key Laboratory of Quantum Information, University of Science and Technology of China, Hefei, 230026, China}
\author{Guang-Can Guo}
\affiliation{CAS Key Laboratory of Quantum Information, University of Science and Technology of China, Hefei, 230026, China}
\affiliation{Synergetic Innovation Center of Quantum Information and Quantum Physics, University of Science and Technology of China, Hefei, Anhui 230026, China}
\affiliation{CAS Center For Excellence in Quantum Information and Quantum Physics,  University of Science and Technology of China, Hefei, Anhui 230026, China}
\author{Ming Gong}
\email{gongm@ustc.edu.cn}
\affiliation{CAS Key Laboratory of Quantum Information, University of Science and Technology of China, Hefei, 230026, China}
\affiliation{Synergetic Innovation Center of Quantum Information and Quantum Physics, University of Science and Technology of China, Hefei, Anhui 230026, China}
\affiliation{CAS Center For Excellence in Quantum Information and Quantum Physics,  University of Science and Technology of China, Hefei, Anhui 230026, China}
\date{\today}

\begin{abstract}
In disordered systems, wave functions in the Schr\"{o}dinger equation may exhibit a transition from the extended phase to the localized phase, in which the states at the boundaries or mobility edges may exhibit multifractality. 
Meanwhile, the Critical Phase (CP), where all states exhibit multifractal structures, has also attracted much attention in the past decades. 
However, a generic way to construct the CP on demand still remains elusive. 
Here, a general approach for this phase is presented using two coupled quasiperiodic chains, where the chains are chosen so that before coupling one of them has extended states while the other one has localized states.
We demonstrate the existence of CP in the overlapped spectra in the presence of inter-chain coupling using fractal dimension and minimal scaling index based on multifractal analysis. 
Then we examine the generality of this physics by changing the forms of inter-chain coupling and quasiperiodic potential, where the CP also emerges in the overlapped spectra. 
We account for the emergence of this phase as a result of effective unbounded potential, which yields singular continuous spectra and excludes the extended states in the overlapped regimes. 
Finally, the realization of this CP in the continuous model using ultracold atoms with bichromatic incommensurate optical lattice is also discussed. 
Due to the tunability of the two chains, this work provides a general approach to realizing the CP in a tunable way. 
This approach may have wide applications in the experimental detection of CP and can be generalized to much more intriguing physics in the presence of interaction for the many-body CP.
\end{abstract}

\maketitle


\section{Introduction}\label{sec-introduction}
Decades after its discovery \citep{anderson_absence_1958}, Anderson Localization (AL), inarguable, has remained one of the most active developing fields in physics nowadays \citep{evers_anderson_2008, Lee1985Disordered, Cottier2019Microscopic,
Sierant2020Thouless, Suzuki2021Anderson,lagendijk_fifty_2009, 
roati_anderson_2008,segev_anderson_2013, billy_direct_2008,matis_observation_2022, 
de_luca_anderson_2014,manai_experimental_2015, lugan_anderson_2007,jiang_interplay_2019, 
kawabata_nonunitary_2021,muller_critical_2016}. In the disordered models, the wave functions may be exponentially localized in real space, following the profile that
\begin{equation}
    |\psi({\bf x})|^2 \sim \exp(-2|{\bf x}-{\bf x}_0|/\xi),
    \label{eq-wflocalized}
\end{equation}
with $\xi$ being the localization length. In general, this length can be calculated using the transfer matrix method \cite{Hoffmanbook}.
Following the scaling theory \citep{abrahams_scaling_1979}, it is believed that the eigenstates in one and two-dimensional systems with any uncorrelated disorder and short-range hopping are all localized.
Localization to delocalization transition at finite disorder strength can be found only in three-dimensional systems, where the mobility edges $E_c$ exist. This mobility edge, separating the localized states and extended states, is a phase transition point, at which the localization length scales as 
\begin{equation}
    \xi(E) \sim |E-E_c|^{-\nu},
\end{equation}
at the localized side, with $\nu$ being the critical exponent. 
This localization has significant influences on the function of semiconductor devices in solid materials, thus from its very beginning, great effects have been made to understand the AL in condensed matter physics in the past several decades \citep{evers_anderson_2008}. 
Recently, it has even unlocked a brand new research area in statistical physics in the presence of many-body interaction, that is, the many-body localization \citep{nandkishore_many-body_2015, Dmitry2019Colloquium} 
violating of thermalization, which has greatly broadened our understanding of the non-equilibrium dynamics. 

\begin{figure}[!htbp]
\includegraphics[width=0.45\textwidth]{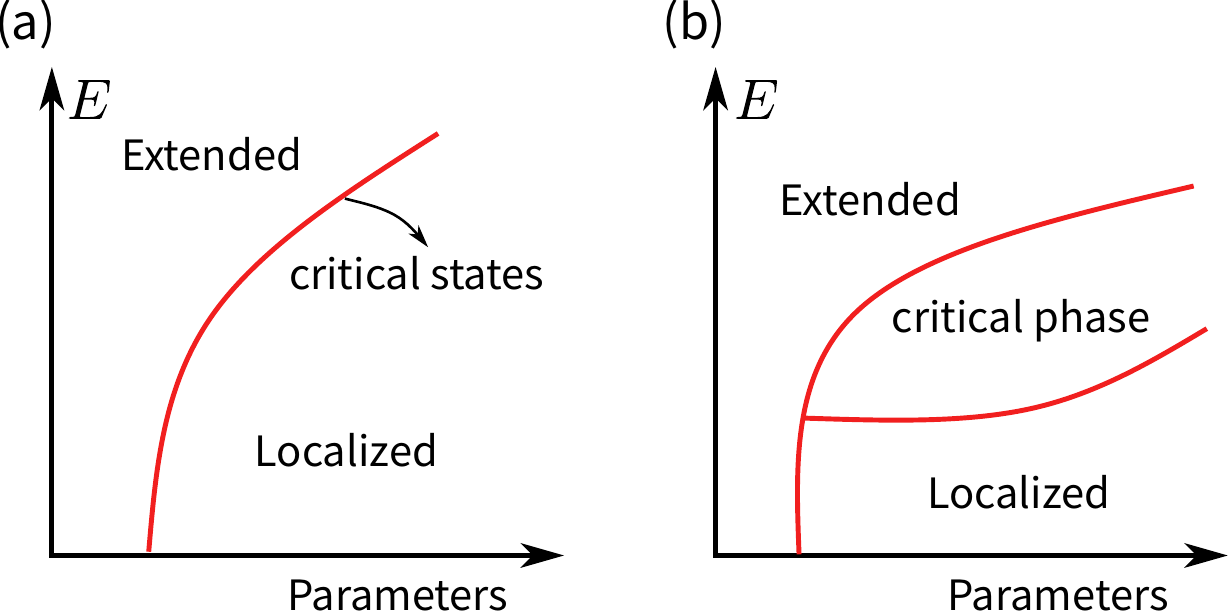}
\caption{
From critical states at the phase boundary (a) to the critical phase in a finite regime (b).}
	\label{fig-critical-phase}
\end{figure}

In these investigations, one of the most active areas is the multifractality of  the critical states during AL transition \citep{janssen_mutifractal_1994, mirlin_statistics_2000,evers_multifractality_2001,rodriguez_multifractal_2008, rodriguez_multifractal_2009,obuse_conformal_2010,
burmistrov_multifractality_2013, pouranvari_multifractality_2019,sbierski_criticality_2021,
miguel_disorder_2020,brillaux_multifractality_2019,
mirlin_multifractality_2000, evers_fluctuations_2000,mirlin_exact_2006,
kravtsov_dynamical_2010,mirlin_multifractality_2000, evers_fluctuations_2000,mirlin_exact_2006,
kravtsov_dynamical_2010,siebesma_multifractal_1987, hiramoto_scaling_1989}.
At the phase boundaries (see Fig. \ref{fig-critical-phase} (a)), these states are neither extended nor localized, exhibiting some totally different features. 
They are extended if we look at them globally, yet they are localized if we look at them locally, that is, they will fluctuate dramatically in real space at all length scales.
Investigation of these states is stimulated by the electronic properties in quasiperiodic systems and the discovery of quasicrystals \cite{Abe1987Fractal}, and sometimes they are termed as intermediate states. 
These states can be characterized using multifractal analysis via the inverse participation ratios
(IPR), or $q$-moment as
\begin{equation}
    P_q = \int d{\bf x} |\psi({\bf x})|^{2q} \sim L^{-\tau_q},
    \label{eq-Pqd}
\end{equation}
where $L$ is the spatial size along one direction.
In general, $\tau_q =0$ is used to characterize the localized state, and $\tau_q = d(q-1)$ ( $d$ is the system dimension) for the extended states. Otherwise, they can be regarded as critical states, with \cite{evers_anderson_2008}
\begin{equation}
    \tau_q = d(q-1) + \Delta_q,
\end{equation}
where $\Delta_q \neq 0$ is the anomalous term. 
Obviously if $q=1$, the wave functions are normalized, with $P_q= 1$; for more details, see Sec. C of Ref. \cite{evers_anderson_2008}. 

\subsection{AL, exact mobility edges, and critical states in 1D quasiperiodic systems}

In the research of the AL, the quasiperiodic potentials have played an unique role, in which the strongly correlated disorder-like potentials make AL transition feasible even in one dimension (1D) models. 
Thus the above physics --AL, mobility edges, and multifractality -- can be researched in 1D systems with quasiperiodic potential. However, the AL transition may also be realized using long-range correlated disorders, which is not the major concern of this work \citep{Shima2004Localization, Moura1998Delocalization}. 
For the physics with quasiperiodic potentials, one of the most prominent examples is the Aubry–Andr\'{e}-Harper (AAH) model \cite{aubry1980analyticity}, which can be written as
\begin{equation}
    H_\text{AAH} = \sum_m Je^{i\phi} b_{m}^\dagger b_{m+1} + \text{h.c.} + 2 V \cos(2\pi \beta m + \phi) b_m^\dagger b_m,
    \label{eq-aah}
\end{equation}
with $b_m$ ($b_m^\dagger$) is the annihilation (creation) operator at site $m$ and $\phi$ is a gauge
phase. This phase will be discussed in subsection \ref{sec-remarks}; otherwise, we will set $\phi=0$.
Throughout
this work, unless defined explicitly, $\beta$ is an irrational parameter, $J$ is the hopping strength and $V$ is the on-site potential depth. A Fourier transformation of the above model via $b_m = 1/\sqrt{N} \sum_k b_k e^{i 2\pi\beta k m}$, yields the following tight-binding model in momentum space
\begin{equation}
    \mathcal{H}_\text{AAH} = \sum_k  2J \cos(2\pi \beta k + \phi) b_{k}^\dagger b_{k} + Ve^{i\phi}b_{k}^\dagger b_{k+1} + \text{h.c.}.
    \label{eq-aah-momentum}
\end{equation}
We find that the above mode in momentum space $\mathcal{H}_{\text{AAH}}$ is exactly the same as that AAH model in Eq. \ref{eq-aah} in real space at the dual point $V = J$, which is independent of the phase $\phi$. 
It is well-known that the Fourier transformation transforms a localized function in real space to an extended function in momentum spcae (e.g, $\delta(x)$ in real space to constant in momentum space), and vice versa.
Therefore, Eq. \ref{eq-aah} and Eq. \ref{eq-aah-momentum} establish a direct mapping between the localized states in one model and the extended states in the other model, which follows the conjecture that the extend to localize transition of Eq. \ref{eq-aah} happens exactly at $V=J$ without mobility edge.
At this point, the AAH model is mapped to itself, leaving the wave functions neither extended nor localized, thus they should be critical states \cite{Abe1987Fractal}.  
The conjecture in the AAH model has been verified by 
numerical calculation \citep{aubry1980analyticity, siebesma1987multifractal}. Here, this transition applies to all states. 
If we only focus on the ground state properties, it can be understood in a much intuitive way. We find that for the extended states, the minimal energy of $H_\text{AAH}$ is $-2J$ for a state $|\psi\rangle =\frac{1}{\sqrt{N}} \sum_m |m\rangle$, while for a fully localized state $|m\rangle$, its minimal energy is $-2V$, thus a crossover between the localized state and extended state happens at $V = J$. 
It is necessary to emphasize that this kind of dual mapping has also played a fundamental role in condensed matter physics, for example, it was used by Kramers and Wannier for the study of the two-dimensional Ising model \cite{Kramers1941Statistics}, in which the high-temperature expansion series is mapped to the low-temperature expansion series. Dual symmetry is also a central concept in condensed matter physics and quantum field theory in some different contexts \cite{Hikida2022Holography, Son2015Composite}.

The idea for searching for mobility edges, based on the dual transformation, has been extended to a broad class of quasiperiodic models. We summarize some of them below. (1) Biddle \textit{et al.} \cite{biddle_predicted_2010} studied a AAH model with long-range hopping as 
\begin{equation}
   H_\text{LAAH} = J\sum_{l\ne m} e^{-p|l-m|} 
   b_l^\dagger b_m + \sum_m V \cos(2\pi \beta m + \delta) b_m^\dagger b_m, 
   \label{eq-LAAH}
\end{equation}
with $p$ a parameter controlling the hopping distance between different sites. When $p$ is small, the long-range hopping is allowed; and when $p$ is large, only the nearest neighboring hopping is important. With a modified Fourier transition, the authors find an exact mobility edge at $\cosh(p) = (E + J)/V$. (2) Ganeshan \textit{et al.} \citep{ganeshan_nearest_2015} develop a nearest neighbor quasiperiodic model as
\begin{equation}
H_\text{GAAH} = \sum_m J(b_m^{\dagger}b_{m+1} + \mathrm{h.c.}) + \frac{2V \cos(2\pi\beta m)}{1 - a\cos(2\pi\beta m)} b_m^{\dagger}b_{m}.
\label{eq-gaah}
\end{equation}
with $|a| < 1$. The exact mobility edge is at $aE = 2\text{sign}(|J|-|V|)$. Recently, this model has been realized in the momentum synthetic lattice of ultracold atomic gas \citep{Alex2021Interaction}, and 
the exact localization length of the GAAH model was proved by the Avila's global theory \citep{Wang2021Duality}. (3) Liu \textit{et al.} \citep{Liu2020GeneralizedAAH} generalized the duality transformation to non-Hermitian systems, where they also found exact mobility edges in the non-Hermitian models.
We emphasize that all those results are possible to be extended to the higher dimensions systems according to the work by Devakul and Huse \citep{Devakul2017Anderson}, in which self-duality can also be realized with these quasiperiodic potentials in the Ando model \cite{Ando1989Numerical}.  

Dual transformation is not the only way to critical states and mobility edges and there are some other ways to find exact phase boundaries of AL in 1D quasiperiodic models.
For instance, Sarma {\it et al. } \citep{Sarma1990Localization} proposed a analytical mobility edge at $E_c = \pm|2-\lambda|$ in a class of model with slow varying potential $V_m = \lambda\cos(\pi \beta m^{\nu})$ with $\nu<1$ using a WKB approximation.
Furthermore, Avila constructed a new theory, now known as Avial's global theory \citep{Avila2015Global}, to exactly prove the Lyapunov exponent $\gamma_{\epsilon}(E)$ of the tight-binding quasiperiodic model is a piecewise convex function of the complex angle $\epsilon$ in $V_m =2V \cos(2\pi\beta m + \theta + \mathrm{i}\epsilon )$.
By using this property, a lot of interesting models without self-duality but with exact mobility edges have been proposed and calculated \citep{wang_one-dimensional_2020, Liu2021Exact, liu2021anomalous, YCZhang2022Lyapunov, Cai2022Exact}.
One interesting example is the Mosaic AAH model \citep{wang_one-dimensional_2020}
\begin{equation}
\begin{aligned}
H_\text{MAAH} &= \sum_m J (b_m^{\dagger}b_{m+1} + \mathrm{h.c.}) \\&
+ 2V(1 + (-1)^m) \cos(2\pi\beta m)b_m^{\dagger}b_{m},
\end{aligned}
\label{eq-mosaic}
\end{equation}
with exact mobility edges at $E_c=\pm J/2V$. While Avila's global theory can be used to determine the extended phase, CP, and localized phase in quasiperiodic models with only nearest-neighbor hopping, it can not be used to study coupled chain models. 

\subsection{From critical states to critical phase}

It has been widely believed that we can find the critical states with multifractality at the phase boundaries of AL. 
In the 3D AL models, these phase boundaries can be determined using numerical methods, while in the above-mentioned 1D models with quasiperiodic potentials, they may be determined analytically.
However, these critical states, even exist, are rare and hardly detected in experiments as they only live at the phase boundaries.
A more intriguing problem for these states is the possible Critical Phase (CP), instead of some critical states.
By CP, here, we mean that all the eigenstates in some parameter regimes, for 
example in a finite energy window, can exhibit multifractality.
With this CP, the experimental detection of these phases may become much easier. 
A recent application of the single-particle CP is the extended but non-thermal many-body CP \citep{wang_realization_2020, cheng_many-body_2021} when the interaction is taken into consideration.
This many-body CP, which interpolates the many-body localization phase and thermalized phase, provides another quantum dynamical phase.

To gain an overview of the CP in quasiperiodic models, here, we summarize some possible candidate models, which have been found in recent years for this phase. Most of these models, as before, involve the quasiperiodic potentials, thus its most general Hamiltonian can be written as
\begin{equation}
\begin{aligned}
    H &= \sum_{l < m} (J_{lm} b_l^\dagger b_m + \Delta_{lm} b_l^\dagger b_m^\dagger + \mathrm{h.c.}) \\
    &+ \sum_m 2V \cos(2\pi \beta m) b_m^\dagger b_m,
\end{aligned}
    \label{eq-LforCP}
\end{equation}
where $J_{lm}$ is the hopping and $\Delta_{lm}$ is the pairing strength between different sites.
(I) In Refs. \cite{wangjun_phase_2016, wang_spectral_2016}, Wang {\it et al.} considered the phase diagram of this model, termed as non-abelian AAH model with $p$-wave pairing, assuming only nearest neighboring hopping and pairing, with $J_{lm} = J_0\delta_{m, l + 1}$ and $\Delta_{lm} =  \Delta \delta_{ m, l + 1}$.
They found the CP locates at $|1-\Delta|<V<1+\Delta$ (for $\Delta >0$), in which all states are critical and exhibit multifractality. 
(II) In Refs. \onlinecite{deng_one-dimensional_2019, roy_fraction_2021}, Deng {\it et al.} and Roy {\it et al.} studied the generalized AAH model with power-law hopping that $J_{lm} = J_0/|l-m|^p$ and without pairing ($\Delta_{lm} =0$).
They found that the CP can even exist at the limit $V \rightarrow\infty$ when $p<1$.
(III) In Ref. \onlinecite{Takada2004Statistics}, Takada \textit{et al.} studied the off-diagonal AAH model with hopping $J_{lm} = (1 + \lambda\cos(2\pi \beta l + \phi))\delta_{m, l + 1}$ and without pairing ($\Delta_{lm} =0$). They found that the CP exists at $|\lambda|>\max(1, V)$. Meanwhile, there are some other physical models for the CP that are related to the unbounded potential. For example,  Liu \textit{et al.} \citep{liu2021anomalous} and Zhang \textit{et al.} \citep{YCZhang2022Lyapunov} demonstrated that the existence of CP for the unbounded potential in Eq. \ref{eq-LAAH} with $a>1$ using multifractal analysis and Avila's global theory.
These phases are much easier to be experimental verified in the current experiments using ultracold atoms and photonic crystals \citep{schreiber_observation_2015, Kohlert2019Observation, Goblot2020Emergence}.

Although there is already a lot of research on the CP in quasiperiodic systems, the underlying mechanism for its emergence is still unclear, and their experimental realization is also of great challenge. An even more challenging problem is the tunability of the CP on demand in experiments. 
To this end, in this work, we provide an alternative general approach for generating of the CP for the above problems.
Unlike the previous research, where the CP was studied case by case, our approach has some kind of generality and can be applied to a wide class of quasiperiodic models, in which the positions, and structures of the CP can be engineered on demand.
Our key idea is to use two coupled quasiperiodic chains, which are chosen in such a way that before coupling one of them has localized states and the other chain has extended states, then the direct coupling between these two chains can be used to construct the CP in their overlapped spectra.
By using FD and minimal scaling index based on multifractal analysis, we demonstrate their multifractality
and examine its generality in several forms of inter-chain couplings and quasiperiodic models.
Our approach should greatly broaden the understanding of CP and shed light on much more intriguing physics based on these CPs.

\subsection{Organization of this work}

This work is organized as follows. 
In Sec. \ref{sec-approach}, we introduce our generic approach to CP. In Sec. \ref{sec-minimalmodel}, a minimal model for the CP is presented, and we briefly review the methods of multifractal analysis, including fractal dimension and scaling index, 
which can be used to identify the critical, localized, and extended states, respectively.
With these methods, we demonstrate that in the overlapped spectra regime all states are critical, while in the un-overlapped regimes, they are either localized or extended.
In Sec. \ref{sec-general}, we examine the generality of this approach to CPs by changing the forms of inter-chain coupling and quasiperiodic potential.
Then we propose a coupled dual quasiperiodic model, in which all states are critical at the self-dual condition. 
Wave packet dynamics in this condition will exhibit sub-diffusion law.
In Sec. \ref{sec-mechansim}, we explain 
the reason why states in the un-overlapped spectra maintain extended or localized while the states in the overlapped spectra become multifractal. We then related them to the applicability of perturbation theory and effective unbounded potential.
In Sec. \ref{sec-observation-realization}, we discuss the possible observation of CPs in a continuous spin-dependent bichromatic lattice with ultracold atoms. Finally, we conclude in Sec. \ref{sec-conclusion} and discuss the significance of this idea for possible new physics. In Appendix \ref{sec-appendix-soc} and \ref{sec-appendix-experiment}, we have also discussed the CP in a spin-orbit coupled lattice model and the realization of spin-dependent bichromatic incommensurate lattice using ultracold atoms. In Appendix 
\ref{sec-appendix-wavepacket}, we show the dynamics of the wave packet diffusion of the coupled dual chains at other parameters.

\begin{figure}[!htbp]
\includegraphics[width=0.45\textwidth]{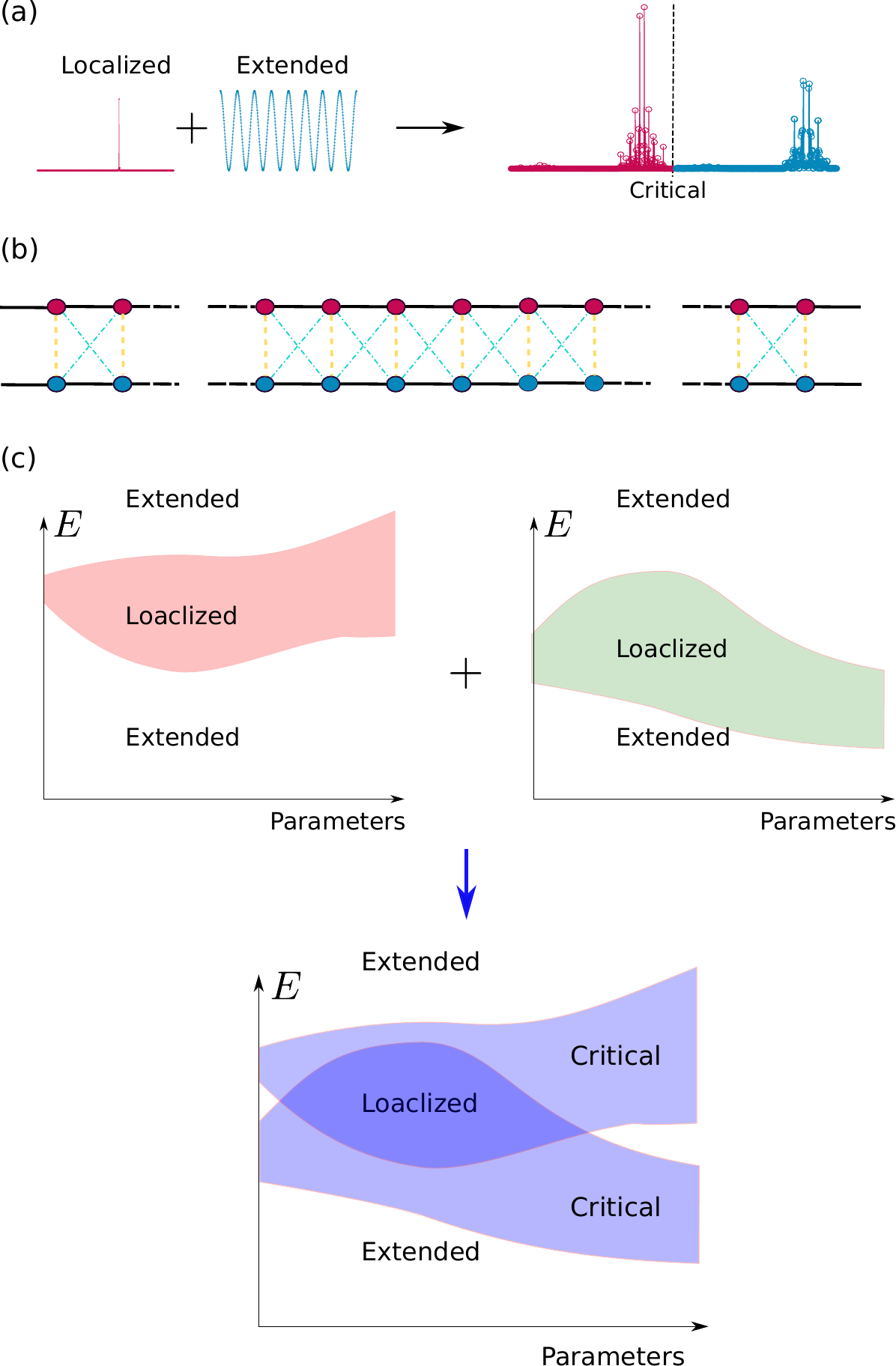}
\caption{The general approach to CP.
(a) States with large spatial fluctuation can be naturally realized by coupling between localized and extended states.
(b) Model for CP based on two coupled quasiperiodic chains, one of which has extended states and the other has localized states when the inter-chain coupling is switched off.
(c) The CP may be found in their overlapped regime when $t_\text{v}\neq 0$ and $L\rightarrow \infty$. 
The upper row represents the phase of one quasiperiodic chain while the below row represents the phase of the coupled quasiperiodic chains.
In this case, the coupling between the same phase will not change the nature of this phase; yet the coupling between the localized phase in one chain and the extended phase in the other chain can yield the critical phase (CP) in the overlapped spectra.}
\label{fig-approach}
\end{figure}

\section{The general approach to the CP}\label{sec-approach}

Firstly, it is crucial to clarify our generic approach to the CP. The system we considered is made of two 1D chains (see Fig. \ref{fig-approach} (b)) with quasiperiodic potentials and inter-chain coupling. 
The inter-chain coupling is chosen to be local in spatial space, while its specific form is not essential. The general 
Hamiltonian reads as    
\begin{equation}
H = H_{1} + H_{2} + H_\text{c},
\label{eq-H-general}
\end{equation}  
where $H_1$ and $H_2$ represent the Hamiltonian for the two chains with quasiperiodic potentials, and $H_c$ is the inter-chain coupling between them. 
We demonstrate that the CP can be realized in the overlapped spectra of the localized phase and extended phase (see Fig. \ref{fig-approach} (c)).  
It means that if one of the chains has extended states and the other chain has localized states, then the inter-chain coupling induces reconstruction of the wave functions in the overlapped spectra regime due to the direct hybridization between them.
These states will exhibit multifractality. 
 
This idea comes from three intuitive motivations. Firstly, the states in the CP have zero Lyapunov exponents with large spatial fluctuation at all length scales, which are recognized as extended states globally, yet localized states locally in some sense \cite{hiramoto_scaling_1989}.  Secondly, this coupling should not yield the coexistence of localized states and extended states, otherwise, an infinite number of mobility edges emerge, making the concept of mobility edges to be ill-defined.
Thirdly, the large spatial 
fluctuation is a natural consequence of coupling between localized states and extended states; see Fig. \ref{fig-approach} (a).
From the perspective of extended states, the effect of localized states causes some kind of random potential, yielding localization effect. However, from the perspective of localized states, the coupling with the extended states can cause delocalization. The competition between these two mechanisms takes responsibility for the emergence of CP in the quasiperiodic potential. 

Without loss of generality, we choose the form of $H_1$ and $H_2$ as the tight-binding model with nearest-neighbor hopping and quasiperiodic potential, with Hamiltonians as
\begin{eqnarray}
H_{1} &&= \sum_m J_{1} ( b_m^{\dagger}b_{m+1} + \mathrm{h.c.}) + g_{1}(m) b_m^{\dagger}b_{m}, \label{eq-cqpc-a} \\
H_{2} &&= \sum_m J_{2} ( a_m^{\dagger}a_{m+1} + \mathrm{h.c.}) + g_{2}(m) a_m^{\dagger}a_{m}, \label{eq-cqpc-b} 
\end{eqnarray}
Here $g_i(x)$ is  a quasiperiodic function for $i=1$, $2$ (e.g. $g(m) = 2 V\cos(2\pi \beta m)$ for the AAH model, $g(m) = 2 V \cos(2\pi \beta m)/(1 - a \cos(2\pi \beta m))$ for the GAAH model, $g(m) = 2V(1 + (-1)^m)\cos(2\pi \beta m)$ for Mosaic AAH model discussed in the introduction). The simplest inter-chain coupling may take the following form
\begin{equation}
H_c = t_\text{v} \sum_m a_m^\dagger b_m + \text{h.c.},
\end{equation}
where $t_\text{v}$ is the coupling strength between the two chains. We will focus on the weak coupling regime, thus 
$|t_\text{v}| \ll |J_1|, |J_2|$. Yet, its form is not essential as long as it is local. In this work, we will discuss other kinds of inter-chain couplings, all of which can give rise to the CP in the overlapped spectra regime, yielding the picture in Fig. \ref{fig-approach} (c). Furthermore, it should be noted that the intra-chain hopping $J_1$ and $J_2$ may include long-range terms (see Eq. \ref{eq-LAAH} and Eq. \ref{eq-LforCP}), which should not affect the main conclusions.  

\section{The minimal model for CP and numerical methods}
\label{sec-minimalmodel}

With the generic approach discussed in the previous section, here we examine the CP in some concrete models. A minimal model, which covers most of the physics of our claims, is to choose $g_2(m) = 0$ and $g_1(m)= 2V \cos(2\pi\beta m)$ the
AAH model. We also set $J_1 = J_2 = 1$ as the overall energy scale. Then we have the following Hamiltonian
\begin{equation}
\begin{aligned}
H &= \sum_m ( b_m^{\dagger}b_{m+1} + \mathrm{h.c.} + 2V \cos(2\pi\beta m) b_m^{\dagger}b_{m}) \\
&+\sum_m ( a_m^{\dagger}a_{m+1} + \mathrm{h.c.}) +  t_\text{v} \sum_m(
a_m^{\dagger}b_m + \mathrm{h.c.}),
\end{aligned}
    \label{eq-minimal-model}
\end{equation}
where $\beta$ is an irrational number. 
As demonstrated by different methods, the AAH chain ($b$ operator in Eq. \ref{eq-minimal-model}) has an Anderson transition without mobility edge at the self-dual point $V=1$ (when $t_\text{v} =0$).
Thus all the states in the AAH chain become localized (extended) at $V>1$ ($V<1$). On the other hand, the purely extended chain ($a$ operator in Eq. \ref{eq-minimal-model}) has energy spectra in the energy window $|E|<2$.
This energy window can be changed by a possible shift of chemical potential, say $\mu \sum_m a_m^\dagger a_m$, which can tune the position of the overlap spectra between the two coupled chains. 
Thus, when $V<1$, the inter-chain coupling in Eq. (\ref{eq-minimal-model}) hybridizes extended states with the two chains in the two chains, yielding extended states. 
However, the physics is totally different when $V>1$, where hybridization between localized states and extended states can yield CP in the overlapped spectra regime (see Fig. \ref{fig-approach} (b) and (c)).

We apply the exact diagonalization method to calculate the eigenstates and eigenvectors of $H$, and then perform a finite-size analysis to extrapolate the physics in the thermodynamic limit. 
The irrational number is chosen as $\beta=\frac{\sqrt{5}-1}{2}$, which can be approximated by the ratio of two Fibonacci numbers $F_{n-1}$ and $F_n$ as $\beta_n=F_{n-1}/F_{n}$, with $\lim_{n\rightarrow} \beta_n = \beta$.
We have used periodic boundary condition in our numerical simulation, assuming wave function as $|\psi\rangle = \sum_{m} \psi_{m,1} b_m^\dagger|0\rangle + \psi_{m,2} a_m^\dagger|0\rangle$, where $|0\rangle$ is the vacuum state.
The localization properties of the eigenvectors are identified by the IPR as 
\begin{equation}
    \mathrm{IPR} = \frac{\sum_{m,\mu} |\psi_{m,\mu}|^4}{(\sum_{m,\mu} |\psi_{m,\mu}|^2)^2},
    \label{eq-ipr}
\end{equation}
and normalized participation ratio (NPR)
\begin{equation}
    \mathrm{NPR} = \frac{(\sum_{m,\mu} |\psi_{m,\mu}|^2)^2}{(L \sum_{m,\mu} |\psi_{m,\mu}|^4)},
    \label{eq-npr}
\end{equation}
with $L$ being the number of sites. Therefore, $\text{IPR} \cdot \text{NPR} = 1/L$ for a specific state. The IPR (NPR) approaches zero (finite) for extended states and critical states, and finite (zero) for localized states \cite{li_mobility_2017} at the limit of $L\rightarrow\infty$.
To identify the critical states from the extended states, one needs to view the scaling of the IPR on the size of the system,
which gives $\text{IPR} \propto L^{-\tau}$. The index $\tau = 1$ for extended states while $0< \tau < 1$ for the critical states. 

It is necessary to remark that the IPR is well-defined. For the localized phase, the IPR will approach a finite value of the order of $\mathcal{O}(1)$. For the extended phase, one may assume the wave function using a plane-wave basis as 
\begin{equation}
    \psi(x) = \sum_n c_j \sqrt{\frac{2}{L}} \sin(\frac{j\pi x}{L}),
\end{equation}
where $c_j$ are constants independent of system length $L$, with $\sum_j |c_j|^2 = 1$. In principle, this wave function can exhibit some short-wave spatial fluctuation, yet it is not a critical state. If we calculate the IPR, we can immediately find that 
\begin{equation}
    \text{IPR} \propto L^{-1},
\end{equation}
which has the same scaling behavior as a single-plane wave. Thus, for a critical phase, we require that $c_j$ also depend strongly on the system size, yielding some more different scaling exponents. By definitions in Eq. \ref{eq-ipr} and Eq. \ref{eq-npr}, we have NPR approaches a constant in the extended phase, and NPR $\sim 1/L$ in the localized phase. For the critical phase, we therefore expect
\begin{equation}
\text{IPR} \sim L^{-\tau}, \quad 
\text{NPR} \sim L^{\tau-1},
\end{equation}
with exponent $\tau \in (0, 1)$. 
To get this exponent, one needs to apply the multifractal analysis, which will be briefly reviewed in Sec. \ref{sec-multifract}, and more details can be found in Refs. \citep{evers_anderson_2008, hiramoto_scaling_1989, wang_spectral_2016}.

\subsection{Multifractal analysis of the wave functions: fractal dimension and scaling index}
\label{sec-multifract}

Multifractality was first introduced by Mandelbrot to characterize the self-similar cascades in turbulence \cite{Mandelbrot1974Intermittent}, which is also a typical feature of wave functions during phase transition. Following Eq. \ref{eq-Pqd}, we define the multifractality of the normalized eigenstate using IPR as \cite{evers_anderson_2008, evers_fluctuations_2000, miguel_disorder_2020, hiramoto_scaling_1989} 
\begin{equation}
    P_q = \sum_{m,\mu} |\psi_{m,\mu}|^{2q},
\end{equation}
with $q=2$ the definition of IPR. Obviously when $q=1$, $P_q = 1$ for a normalized state. When $q =1+\epsilon$, $(P_{0}- P_q)/\epsilon$ will directly gives the on-site entropy. In general, the $P_q$ exhibits a power law dependence on the system size as 
\begin{equation}
    P_q \propto L^{-\tau_q} = L^{-D_q(q-1)}.
\end{equation}
The $D_q$ is the so-called fractal dimension (FD) of the wave function. In general, we may write $\tau_q = (q-1) + \Delta_q$ \cite{Jagannathan2021Fibonacci}, thus $\Delta_q$ is the anomalous term.
For a localized state, we naturally have $D_q = 0$, and $P_q$ approaches a constant. For an extended state, we may simply assume $|\psi| = 1/\sqrt{L}$ \footnote{In $d$-dimensional system, we 
should have $\tau_q = d(q-1)$, with a uniform wave function $|\psi| = 1/L^{d/2}$. In this work, we focus only on $d=1$. } for an uniform distribution (in 1D) and we should have 
\begin{equation}
    P_q \propto L^{-(q-1)}.
\end{equation}
Thus $D_q = 1$ and 0 for extended and localized states, respectively, in the thermodynamic limit $L\rightarrow\infty$. However, the critical states exhibit large spatial fluctuation, which makes $D_q$ to be dependent on $q$ and $0<D_q<1$ in the large $L$ limit. In our finite size calculation, we define the finite size FD as 
\begin{equation}
D_q(L)= -\frac{\log(P_q)}{\log(L)(q-1)}.
\end{equation}
Naively, in the large $L$ limit, we expect that  \begin{equation}
     D_q(L) = D_q + \mathcal{A}(L,q)/\log(L).
\end{equation}
with $\mathcal{A}(L,q)$ ato be a coefficient. In this work, unless defined explicitly, we set $q=2$, with 
$D_2(L) = \tau_2(L)$. Thus, we call $\tau_2(L)$ the wave function FD in this manuscript. 

Another complementary quantity to characterize the multifractal structure of the wave function is the scaling index $\alpha$ \citep{wangjun_phase_2016, evers_anderson_2008}, defined based on the on-site magnitude of the eigenstate
\begin{equation}
    |\psi_{m, \mu}|^2 = \mathcal{A}^2 L^{-\alpha_{m\mu}(L)}\propto L^{-\alpha_{m\mu}(L)},
\end{equation}
with $|\psi_{m,\mu}|^2$ the amplitude on $m$-th site and $\mu$-th chain, and $\mathcal{A}$ is the normalized constant. 
With the distribution of $\alpha_{m\mu}$, we can define $\Omega_L(\alpha)d\alpha$ as the number of lattice sites in a finite interval $\alpha$ and $\alpha +d\alpha$. The distribution function $\Omega_L(\alpha)$ 
should have lattice size dependence, from which we can define the singular spectrum $f_L(\alpha)$ as
\begin{equation}
    \Omega_L(\alpha) \propto L^{f_L(\alpha)}
\end{equation}
The singular spectrum in the large $L$ limit $f(\alpha)$ can be calculated as $f(\alpha) =\mathrm{lim}_{L\rightarrow\infty} f_L(\alpha)$.
For the extended wave functions, all the lattice sites almost have the same 
amplitude $|\psi_{m,\mu}|^2 \sim 1/L $; thus $f(\alpha)$ is only defined at $\alpha = 1$ with $f(\alpha)=1$. 
For localized wave functions, on the other hand, $|\psi_{m,\mu}|^2 $ is nonzero only on a finite number of lattice sites. These sites have an index $\alpha=0$, and the remaining sites with an exponentially small probability measure have $\alpha=\infty$; thus $f(\alpha)$ takes only two values: $f(\alpha=0) = 0$ and $f(\alpha = \infty) = 1$.
For critical wave functions, the index $\alpha$ has a distribution and hence the singular spectrum $f(\alpha)$ is a smooth function defined on a finite interval $[\alpha_{\mathrm{min}},\alpha_{\mathrm{max}}]$.
Therefore, to identify the extended, critical, and localized wave functions, we may simply examine the minimum value of the index $\alpha_{\mathrm{min}}$ for a chosen wave function, which should take 
\begin{equation}
\alpha_{\mathrm{min}}=
    \begin{cases}
        1, & \quad \text{Extended } \\
        (0, 1), &\quad \text{Critical} \\ 
        0, & \quad \text{Localized.}
    \end{cases}
\end{equation}
in the large $n$ limit.
In the numerical calculation, we follow the method by Hiramoto and Kohmoto \citep{hiramoto_scaling_1989}. The finite size scaling of $\alpha_{\mathrm{min}}$ can be written as
\begin{equation}
    \alpha_{\min}(n) = \alpha_{\min}(n\rightarrow\infty) + \frac{\mathcal{A}(n)}{n}, 
\end{equation}
with $n$ is the Fibonacci index and $\mathcal{A}(n)$ is some fitting constant. The number of lattice sites is given by $L = 2F_n$ with $F_n$ the $n$-th Fibonacci number. In this work, the largest calculated system size with exact diagonalization method is $n=20$, and $L=2F_{20} = 13530$, with $1/\ln(L) = 0.10512$. 

\begin{figure}[!htbp]
\includegraphics[width=0.45\textwidth]{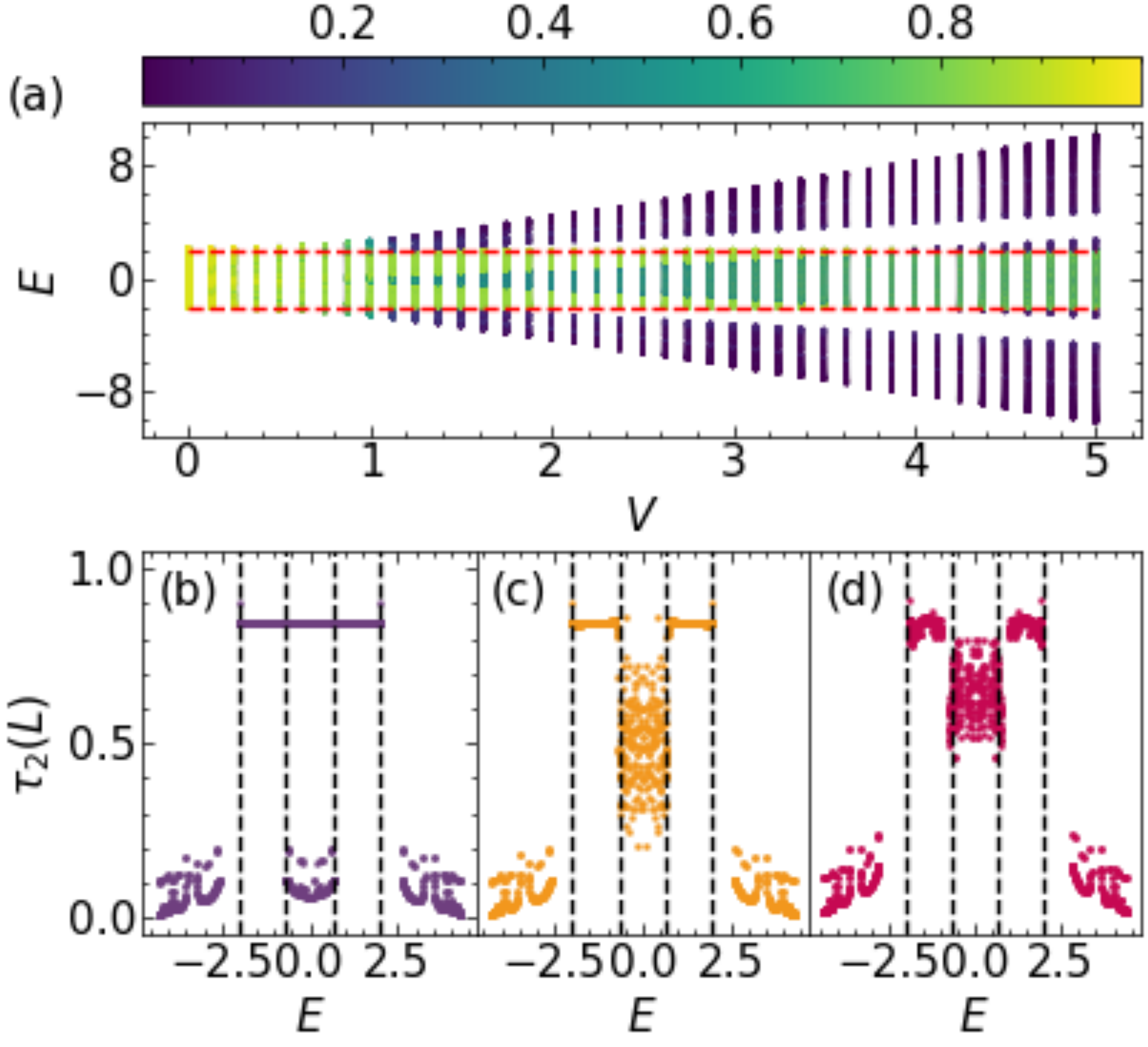}
\caption{(a) FD $\tau_2(L)$ of all wave functions of the minimal model in Eq. \ref{eq-minimal-model} with $\alpha = F_{14}/F_{15} = 377/610$, $t_\text{v}=0.1$ versus potential depth $V$. 
The red dashed lines denote the energy $E=\pm 2$.
(b)-(d) $\tau_2(L)$ versus energy $E$ with $V=2$ and $t_\text{v} = 0$ (left), $0.1$ (middle) and $0.5$ (right). The mobility edges are given by the vertical dashed lines at $E = \pm 2$, and $\pm 0.67$.}
\label{fig-minimal}
\end{figure}

\begin{figure}[!htbp]
\includegraphics[width=0.45\textwidth]{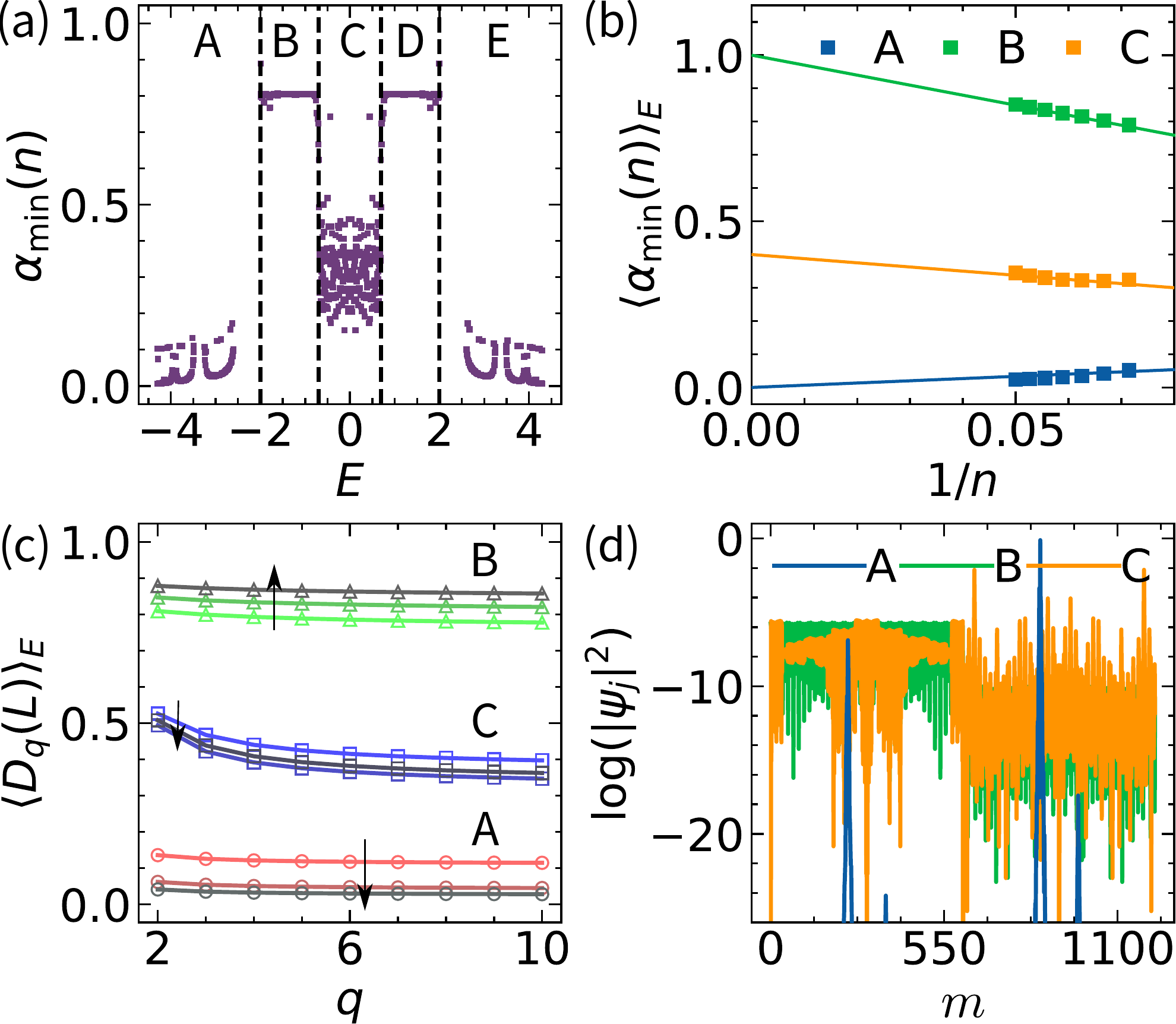}
\caption{(a) Minimal scaling index $\alpha_{\mathrm{min}}$ of all wave functions of the minimal model in Eq. \ref{eq-minimal-model} 
with $\beta = F_{17}/F_{18} = 1597/2584$, $L=5168$, $t_\text{v} = 0.1$, and $V = 2$. 
Those mobility edges are the same as that in Fig. \ref{fig-minimal} (c) with $E = \pm 2$, $\pm 0.67$.
(b) Scaling of the average $\langle \alpha_{\mathrm{min}}(n)\rangle_E $ versus the inverse of Fibonacci index $n$ of the states in energy regimes A, B, and C as panel (a). 
The $n$ is ranging from $14$ to $20$ with the maximum system size is $L=2F_{20}=13530$.
The solid lines serve as a guide for the eyes. 
(c) The averaged fractal dimension $\langle D_{q}(L)\rangle_E$ in different energy regimes A, B, and C as panel (a). We use $n=12$, $n=15$, and $n=19$ in this panel and label their increasing direction as the arrows. 
(d) $\log|\psi_{m}|^2$ versus lattice site $m$ for eigenstate $\psi_{100}$ in A, $\psi_{305}$ in B, $\psi_{610}$ in C with $L = 2F_{15} = 1220$. 
The site index of $m < L/2$ for the free chain and $ m \ge L/2$ for the AAH chain.}
\label{fig-scaling}
\end{figure}

\begin{figure*}[!htbp]
\includegraphics[width=0.8\textwidth]{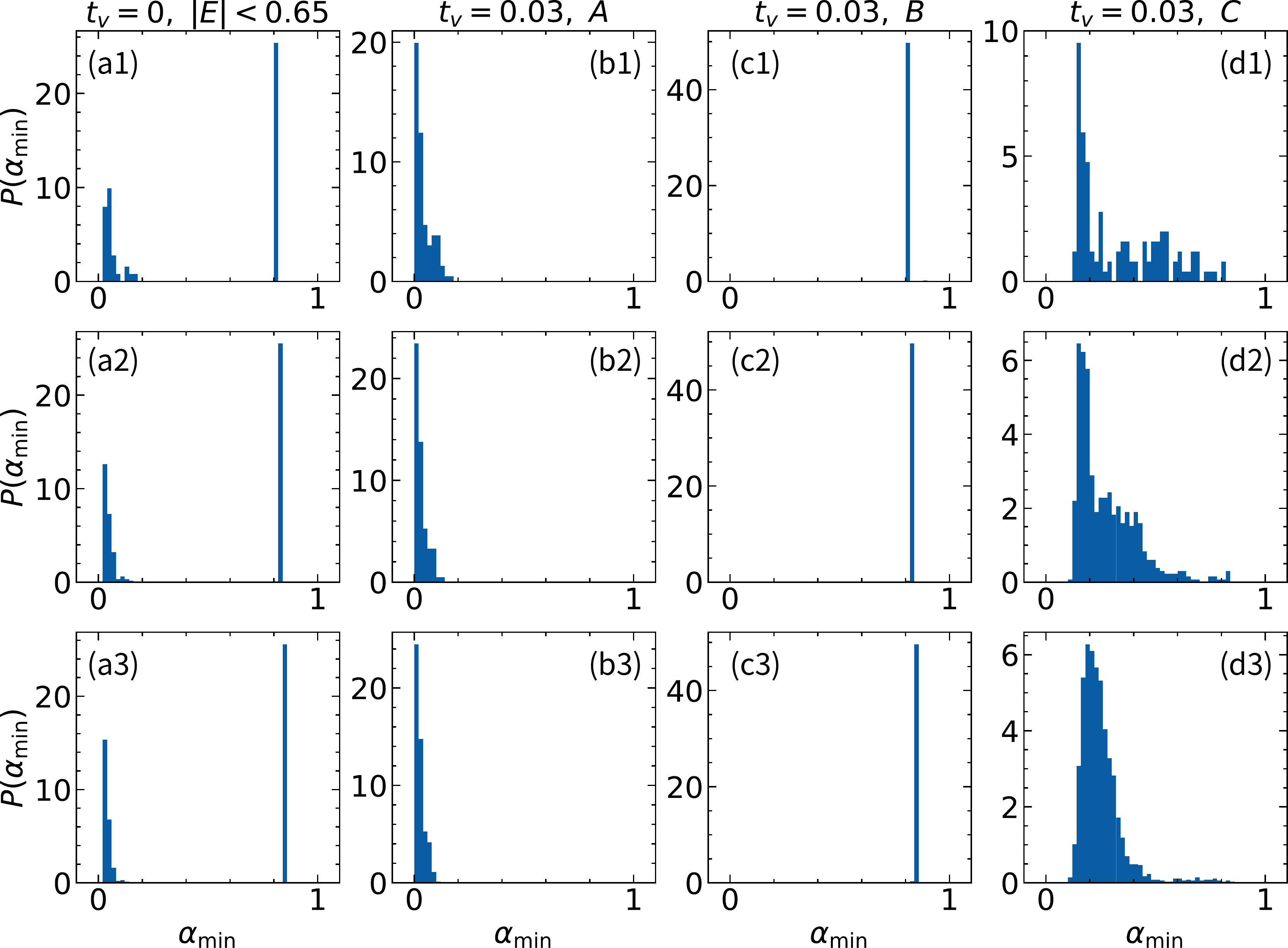}
\caption{
 Distribution of $\alpha_{\mathrm{min}}$ in different phases with the increasing of lattice size (top row $n = 15$; middle row $n = 17$; bottom row $n = 19$). The potential strength is set to $V=2$, the same as in Fig. \ref{fig-scaling}.
(a1)-(a3) The states in the coexistence phase with $t_\text{v}=0$ and $|E|<0.67$, in which $P(\alpha_{\mathrm{min}})$ exhibits a bimodal structure independent of system sizes. 
$P(\alpha_{\mathrm{min}})$ of the states in the energy regime A (see (b1)-(b3)), and energy regime B (see (c1)-(c3). 
(d1)-(d3) $P(\alpha_{\mathrm{min}})$ of the states in the energy regimes C with a wider distribution.
The energy regimes A, B, and C correspond to that in Fig. \ref{fig-scaling}.
}
\label{fig-distribution}
\end{figure*}

\subsection{On the difficulties of the iterative transfer matrix method for quasiperiodic models}

Since the states in the extended phase, localized phase, and CP have different behavior for quasiperiodic systems \citep{Avila2015Global}, one may expect the identification of these phases using the iterative transfer matrix method \cite{Hoffmanbook}, which has been one of the most important tools in the determination of localization length.   
Although this method is efficient for the random potential models, it faces some difficulties for the quasiperiodic model as we explained below. 
In the traditional iterative transfer matrix method, we first select the energy $E$ of the transfer matrix and then obtain the Lyapunov exponent $\gamma(E) = \xi(E)^{-1}$ after performing iterations.
It's well-known that the $\gamma(E)$ measures the asymptotic growth/decay rate of the wave functions when $E$ is inside the spectrum, while it does not have any physical meaning when $E$ is outside the spectrum.
Thus, to research the physics of AL, we need to choose $E$ which is located in the spectrum.
This task can be done well for systems with random potential. 
However, determining whether $E$ is inside the spectrum is a very tough task for quasiperiodic systems, especially for the CP.
This can be understood from the blowing facts.
For random potential models, their energy level spacings are proportion to $L^{-1}$ and their energy levels are densely distributed (without any minimal gap) in the large $L$ limit. 
In contrast, there are infinite numbers of small gaps (with self-similar structures) for the quasiperiodic models in the limit $L\rightarrow \infty$ \citep{Hofstadter1976Energy, Diener2001Transition, yao_critical_2019}, which call for more careful treatments on $E$.
The usual treatment to get $E$ is to use the eigenvalue obtained from a finite-size system.
However, it's wondered that much of the energy $E$ which is chosen in this way is more likely located in the gaps since the self-similar structure of the spectrum.
This should account for the significant fluctuation of localization length $\xi(E)$ against energy $E$ in the numerical calculation of quasiperiodic systems.
Another treatment is to calculate the density of states of the quasiperiodic models in the large $L$ limit, which is still an open question.
The more severe thing is that the spectrum of the CP in quasiperiodic models is singular continuous with zero measures \citep{Hofstadter1976Energy, Avila2015Global}, which means the density of states is not even well-defined.
We emphasize that similar results (large fluctuations for the $\gamma(E)$) were also reported by Li \textit{et al.} \citep{li_mobility_2020} and Zhang \textit{et al.} \citep{YCZhang2022Lyapunov}, where the Lyapunov exponent $\gamma(E)$ from numerical method does not agree with the analytical expression.
Therefore, we conclude the traditional iterative transfer matrix method is not numerically reliable for the CP of quasiperiodic models.
In literature for AL with quasiperiodic potentials, it is better to analyze the structure of the wave functions to fully determine their multifractality in finite-size systems.
Then we should perform careful finite-size scaling to extrapolate the physics in the limit $L\rightarrow\infty$. 
This is the underlying logic of subsection \ref{sec-multifract}.

\subsection{Critical states in the overlapped spectra}

By using FD and minimal scaling index, we can completely characterize and distinguish the critical states from the localized and extended states. 
In order to make sure the irrational limit of $\beta = \frac{\sqrt{5}-1}{2}$ is well approximated, we enlarge the size of the system using the Fibonacci sequence with the total number of sites as $L = 2F_n$ and $\beta_n = F_{n-1}/F_n$.
The FD $\tau$ as a function of $V$ is presented in Fig. \ref{fig-minimal} (a). When $0 < V <1$, all states are extended in both chains, $\tau> 0.8$ is realized. When $V>1$, localization happens in one of the chains, and the critical states emerge in the overlapped spectra with $0.2<\tau_2(L)<0.8$.
The extended and localized are found in the un-overlapped spectra regime. This picture is the same as 
Fig. \ref{fig-approach} (c). 

To verify the effect of inter-chain coupling, we present the results for  $t_\text{v}=0$, $t_\text{v}=0.1$, and $t_\text{v}=0.5$ with $V=2$ in Fig. \ref{fig-minimal} (b)-(d).
When $t_\text{v} = 0$, the two chains are decoupled, so the spectra are made by extended states with $\tau > 0.8$; and localized states with $\tau < 0.2$. The finite size scaling in Fig. \ref{fig-distribution} supports our conclusion. 
With finite $t_\text{v}$, we find that the states with $|E| < 0.67$ have $ 0.2 < \tau_2(L) < 0.8$, while the states with $|E| > 2$ ($0.67<|E|<2$) have $ \tau_2(L) < 0.2$ ($\tau_2(L) > 0.8$).
The energy separating the CP and extended phase defines a new kind of mobility edges, whereas the previous research focus on the study of mobility edges between localized states and extended states \cite{biddle_predicted_2010, ganeshan_nearest_2015, li_mobility_2017,
yao_critical_2019,wang_one-dimensional_2020}. 

Since all the wave functions at finite size are critical with FD $0<\tau_2(L)<1$ and $0<\alpha_{\mathrm{min}}(n)<1$, it is extremely important to perform finite-size scaling to understand their large $L$ limits, which are useful for us to identify the critical, localized, and extended states. 
We present the finite-size scaling of these quantities in Fig. \ref{fig-scaling} (b) and (c). Here, we use the averaged value $\langle D_q(L) \rangle_E$ of $D_q(L)$ in an energy interval $[E_{\mathrm{min}}, E_{\mathrm{max}}]$ or $\langle \alpha_{\min}(n) \rangle_E$ for the finite size scaling. 
In the overlapped spectra C, we find $\langle \alpha_{\mathrm{min}}(n)\rangle_E$ approaches 0.4, which is multifractal and belongs to the CP.  In contrast, that in regime A approaches unity, and that in regime B approaches zero when $L\rightarrow\infty$. 
These results demonstrate the states in the overlapped spectra are indeed critical, while in the un-overlapped spectra regimes,
the states are either localized or extended. This result is consistent with the physical picture in Fig. \ref{fig-critical-phase} and 
Fig. \ref{fig-approach}. 

This conclusion is also supported by the behavior of $\langle D_q(L)\rangle_E$ as a function of $L$.
As shown in Fig. \ref{fig-scaling} (c), $\langle D_q(L)\rangle_E$ in regime A (B) approaches unity (zero) when the lattice size is extrapolated to infinity. 
However, $\langle D_q(L)\rangle_E$ approaches a finite value between zero and unity ($\sim 0.5$) with some obvious $q$ dependence. These features suggest that the states in regime C are critical.  We also plot $\log|\psi_{j, \mu}|^2$ versus site index $j$ for three states in Fig. \ref{fig-scaling} (d). 
We see that in regime A, the state exhibits an exponential decay tail, and in regime B it is almost uniformly distributed in the whole chain.
However, in regime C, large spatial fluctuation exists, accounting for the multifractal behavior. 
Similar features are shown in the off-diagonal AAH model \cite{liu_localization_2015} and in higher-dimensional disordered AL models \cite{Hoffmanbook}. 
Thus, we can conclude that the states in regime A are localized, in regime B are extended and in regime C are critical. 

Furthermore, in order to rule out the possibility that $\langle \alpha_{\mathrm{min}}(n)\rangle_E \sim 0.4$ comes from the coexistence of extended states and localized states (such as that for $t_\text{v} = 0$,
in which the averaged $\langle \alpha_{\mathrm{min}}(n) \rangle_E$ is also between 0 and 1), we examine the distribution of $\alpha_{\text{min}}$ and show the results in Fig. \ref{fig-distribution}.
The phase with the coexistence of extended states and localized states would generate two peaks of $P(\alpha_{\text{min}})$. 
However, the CP exhibits a unimodal distribution, indicating that the states in regime C have the same multifractal structure. 

\begin{figure}[!htbp]
\includegraphics[width=0.48\textwidth]{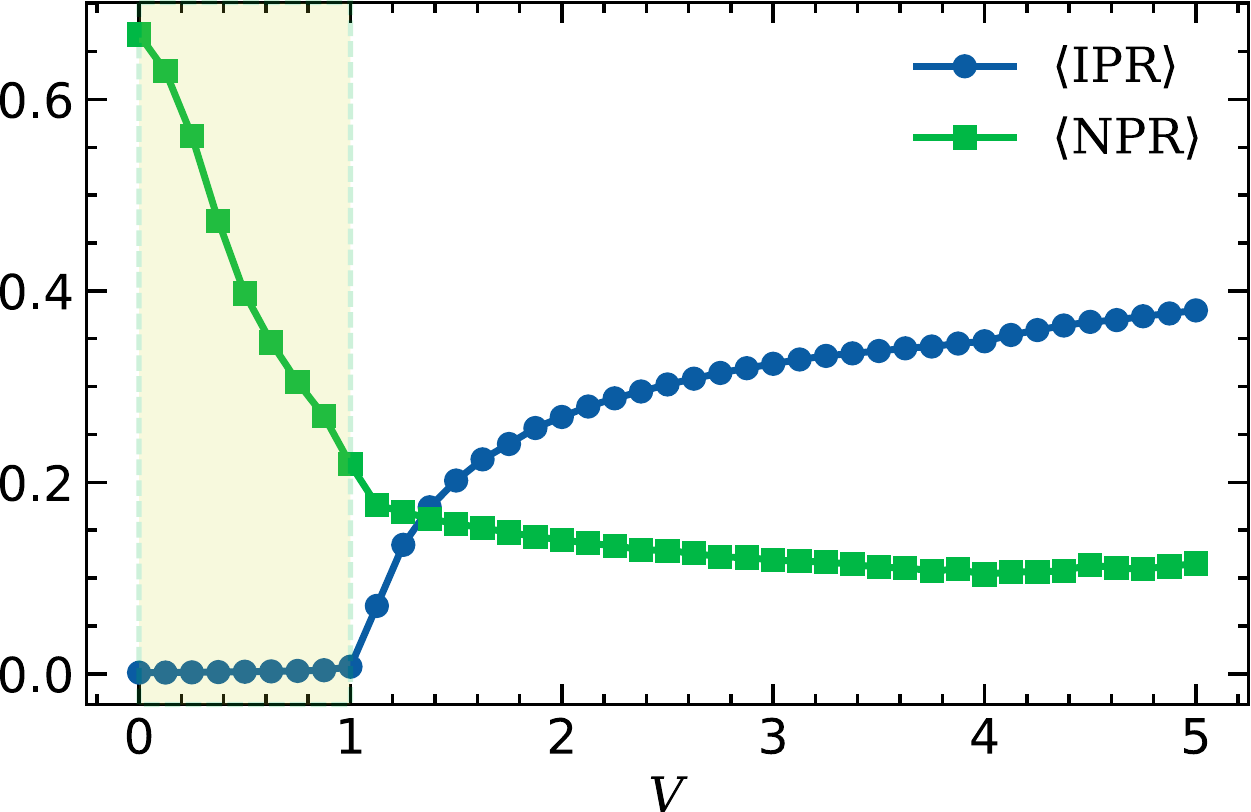}
\caption{The mean IPR and mean NPR of the states in Eq. \ref{eq-minimal-model} versus potential depth $V$. We set $L = 2F_{15} = 1220$, $t_\text{v} = 0.1$, which is same as that in Fig. \ref{fig-minimal}. The yellow shadow represents the extended phase with $V<1$. Similar results are obtained in Fig. 9 of Ref. \onlinecite{li_mobility_2020}. When $V>1$, $\langle \text{IPR}\rangle$ and $\langle \text{NPR}\rangle$ are finite due to the coexistence of extended, localized, and critical phases. }
\label{fig-mean-ipr-npr}
\end{figure}

While these physics can be well discriminated based on the FD and minimal scaling index, it is necessary to emphasize that they can not be determined based on solely mean IPR and mean NPR, which are defined as 
\begin{equation}
\langle \text{IPR} \rangle = \frac{1}{L}\sum_{j=1}^{L} \text{IPR}_j, \quad \langle \text{NPR} \rangle = \frac{1}{L}\sum_{j=1}^{L} \text{NPR}_j.
\end{equation}
While IPR$\cdot$NPR $=1$ for a specific state, their mean will not have this identity. 
We present the results of the mean IPR and mean NPR in Fig. \ref{fig-mean-ipr-npr}, where we find $\langle \text{IPR} \rangle\sim 0$ (finite) at $V<1$ ($V>1$).
The NPR decreases with the increase of $V$, while its value is always finite.
Therefore, we may conclude that all states are extended when $V<1$, which is consistent with our previous analysis.
However, both the value of $\langle \text{IPR} \rangle$ and $\langle \text{NPR} \rangle$ are finite at $V>1$,
which comes from the averaging over localized phase, extended phase, and CP.
Actually, the finite value of $\langle \text{IPR} \rangle$ and $\langle \text{NPR} \rangle$ can also come from averaging over localized phase and extended phase. This feature
can not be changed by the increasing of system size $L$. 
Therefore, to fully characterize the phase of the model in Eq. \ref{eq-minimal-model}, we should use the multifractal analysis, as demonstrated in this section.  In Ref. \citep{li_mobility_2020}, the authors proposed a model with Hamiltonian the same as Eq. \ref{eq-minimal-model} and systematically studied the phase diagram of this model even at a strong $t_\text{v} \gg J$.
They found a large regime, in which the mean IPR and NPR are finite, which is called the single-particle intermediate phase. 
Based on the analysis in this work, we may expect the realization of the CP in this regime with $t_\text{v} \gg J$.
In this regime, the CP would not be limited to the overlapped spectra, because the strong couplings have a great influence on the spectra.
These interesting physics will be left 
for our future investigation. 

\section{The generality of this approach to CP}\label{sec-general}
Although we have already identified the CP in the above minimal model, it is still important to examine the generality of this approach for CP in the other models with quasiperiodic potential. Here, we change the forms of inter-chain coupling and quasiperiodic potentials, showing that all of these models exhibit the same feature for CP. We also present a self-dual
critical model based on two quasiperiodic chains,
in which all states are critical, and the wave packet exhibit sub diffusion dynamics.
All these results demonstrate the generality of our approach for CP, which should have important applications in the research of CP and their associated many-body physics. 

\subsection{The inter-chain coupling}
\begin{figure}[!htbp]
\includegraphics[width=0.48\textwidth]{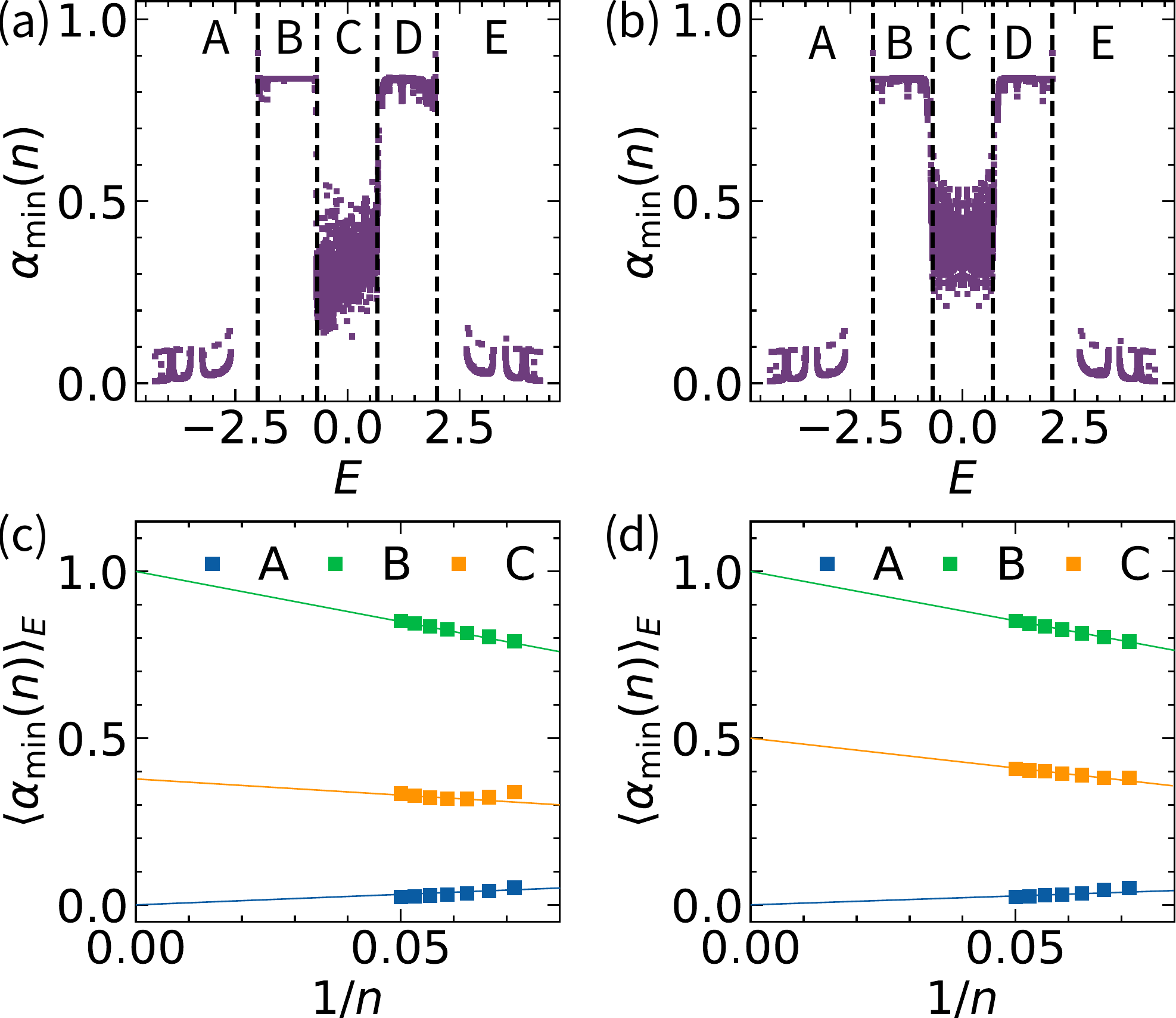}
\caption{The $\alpha_{\mathrm{min}}(n)$ of the all the states in Eq. \ref{eq-couple-nnn} (panel (a)) and Eq. \ref{eq-couple-soc} (panel (b)) for $L = 2F_{18} = 5168$, $t_\text{v}=0.1$, and $V=2$. 
The dashed lines denote the energy boundary $E=\pm 0.67$, $E=\pm 2$. 
The averaged minimal scaling index $\langle \alpha_\text{min}(n)\rangle_E$ in the different energy regimes of the states in with inter-chain coupling in Eq. \ref{eq-couple-nnn} (panel (c)) and in Eq. \ref{eq-couple-soc} (panel (d)) against the inverse of Fibonacci index $n$. The solid lines are guides for the eyes.}
\label{fig-coupling}
\end{figure}

In this section, we discuss two kinds of coupled quasiperiodic chains, where the inter-chain coupling differs from that in Eq. \ref{eq-minimal-model}.
For simplification, we still consider the two chains the same as Eq. \ref{eq-minimal-model} The inter-chain coupling is chosen as 
\begin{equation}
    H_{c} = t_\text{v} \sum_{m} \left(a_m^{\dagger}b_m + a_{m+1}^{\dagger}b_m + a_m^{\dagger}b_{m+1} + \mathrm{h.c.} \right),
    \label{eq-couple-nnn}
\end{equation}
or 
\begin{equation}
    H_{c} = t_\text{v} \left( \sum_{m} (a_{m}^{\dagger}b_{m+1} - a_m^{\dagger}b_{m-1}) + \mathrm{h.c.} \right).
    \label{eq-couple-soc}
\end{equation}
As shown in Fig. \ref{fig-coupling} (a) and Fig. \ref{fig-coupling} (b), the $\alpha_{\mathrm{min}}$ in these two couplings exhibit similar features as that in Fig. \ref{fig-scaling} (a).
The states with $|E|>2$ are localized ($\alpha_\text{min}<0.15$) while the states with $0.67<|E|<2$ are extended (($\alpha_\text{min}>0.8$)).
The states in the overlapped spectra with $|E|<0.67$ are critical ($0.15<\alpha_\text{min}<0.8$).
The finite-size scaling is presented in Fig. \ref{fig-coupling} (c) and Fig. \ref{fig-coupling} (d), which demonstrate our claim. 
Although the phenomenon that the CPs emerge from the overlapped spectra is the same, there are some quantitative different behavior of $\alpha_{\mathrm{min}}$. 
We find $\alpha_\text{min}$ of the states in the model with inter-chain coupling in Eq. \ref{eq-couple-nnn} is asymmetric around $E=0$. However, $\alpha_\text{min}$ in Eq.\ref{eq-couple-soc} is symmetric around $E=0$.

\subsection{The form of quasiperiodic potential}\label{sec-unv-potential}
\begin{figure}[!htbp]
\includegraphics[width=0.48\textwidth]{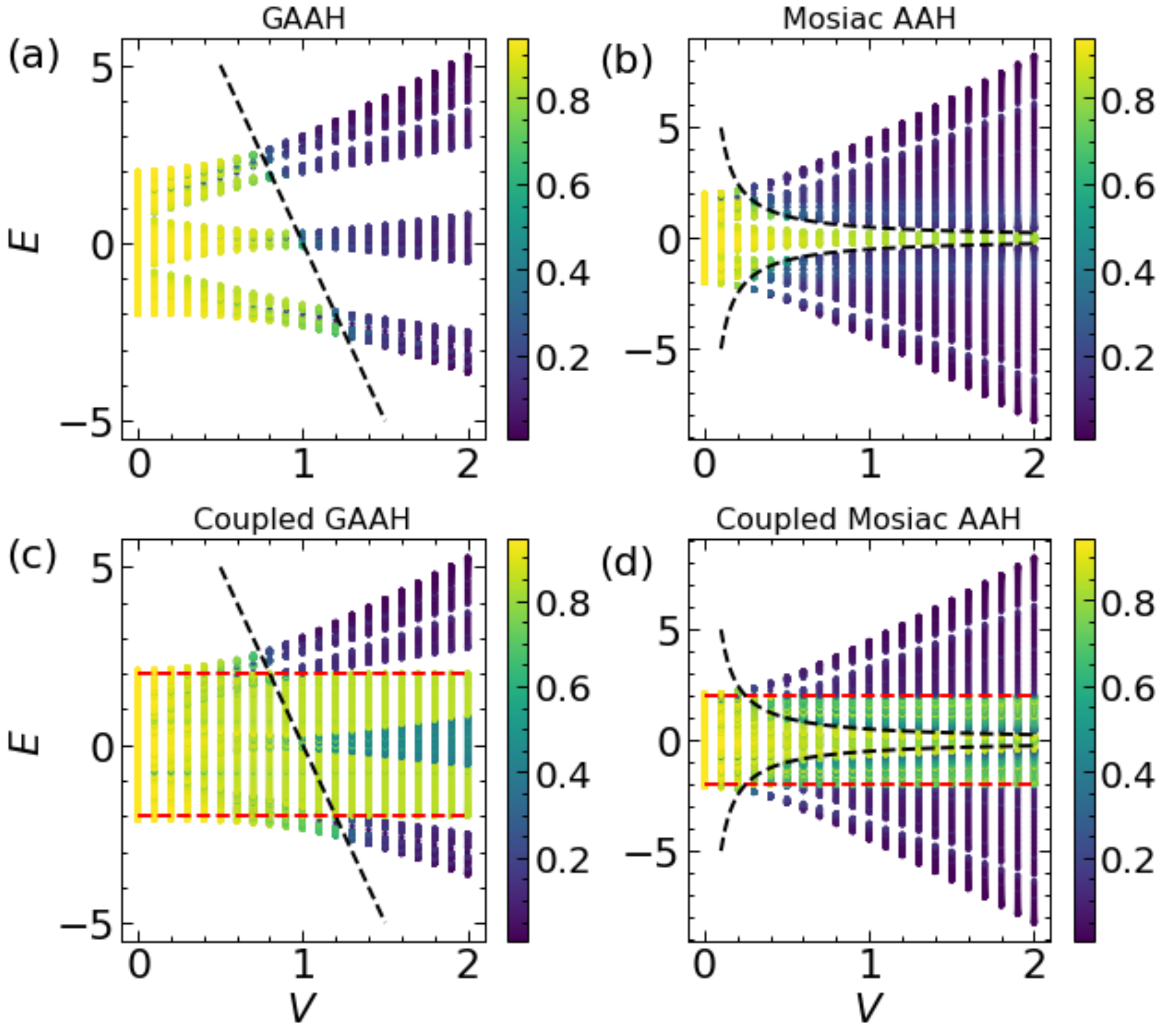}
\caption{
The second FD $\tau_2(L)$ of all states versus $V$ with $t_\text{v}=0.1$ and $L = 2F_{15} = 1220$ for the general AAH model and Mosaic AAH model. 
The first row denotes the mobility edges in the two exact solvable models and the second row denotes the CPs in the coupled quasiperiodic chains. 
The states in the overlapped regimes turn to critical with $\tau \sim 0.25 - 0.75$; see Fig. \ref{fig-potential-detail}.
The black solid lines are the analytical mobility edges and the red solid lines are the energy with $E=\pm 2$. }
\label{fig-potential-overlap}
\end{figure}

To examine the generality of our mechanism for CPs, here we consider two chains with some other types of quasiperiodic potentials, whose mobility edges are analytically well-known. 
The inter-chain coupling is chosen the same as Eq. \ref{eq-minimal-model}. 
The first quasiperiodic chain we choose is the GAAH model. 
The full Hamiltonian of this coupled chain is governed by
\begin{eqnarray}
H &&= \sum_m ( b_m^{\dagger}b_{m+1} + \mathrm{h.c.} + 2V \frac{\cos(2\pi\beta m)}{1-a\cos(2\pi\beta m)} b_m^{\dagger}b_{m}) \nonumber \\
&&+\sum_m ( a_m^{\dagger}a_{m+1} +  t_\text{v}
a_m^{\dagger}b_m + \mathrm{h.c.}),
\label{eq-coupled-gaah}
\end{eqnarray}
According to the constructed dual mapping between localized states and extended states \citep{ganeshan_nearest_2015}, the mobility edge of the GAAH model is given by $a E_\text{c} = 2- 2V$. 
Here, focus on the regime $|a| < 1$ to avoid the unbounded potential, which supports CP.
Therefore, we hope the CPs also emerge above the mobility edges in the overlapped spectra between the localized states in the $b$ mode and extended states in the $a$ mode. 
Another quasiperiodic model we consider is the Mosaic AAH model \citep{wang_one-dimensional_2020}, whose mobility edges are given by $E_\text{c} = \pm \frac{1}{2V}$. The full Hamiltonian of the coupled Mosaic AAH chains is
\begin{equation}
\begin{aligned}
H &= \sum_m ( b_m^{\dagger}b_{m+1} + \mathrm{h.c.} + V_m b_m^{\dagger}b_{m}) \\
&+\sum_m ( a_m^{\dagger}a_{m+1} + \mathrm{h.c.}) +  t_\text{v} \sum_m(
a_m^{\dagger}b_m + \mathrm{h.c.}),
\end{aligned}
    \label{eq-coupled-mosaic}
\end{equation}
with 
\begin{equation}
V_m = 2V((-1)^m+1)\cos(2\pi\beta m).
\end{equation}
The energy spectra and FD of the model in Eq. \ref{eq-coupled-gaah} and in Eq. \ref{eq-coupled-mosaic} are presented in Fig. \ref{fig-potential-overlap} (a) and (b), where the FD calculations fit well with the analytical mobility edges.
The FD of the states in coupled chains against the disorder strength $V$ is shown in Fig. \ref{fig-potential-overlap}, in which the critical states can be realized in the overlapped spectra regime in these two different models. 

\begin{figure}[!htbp]
\includegraphics[width=0.48\textwidth]{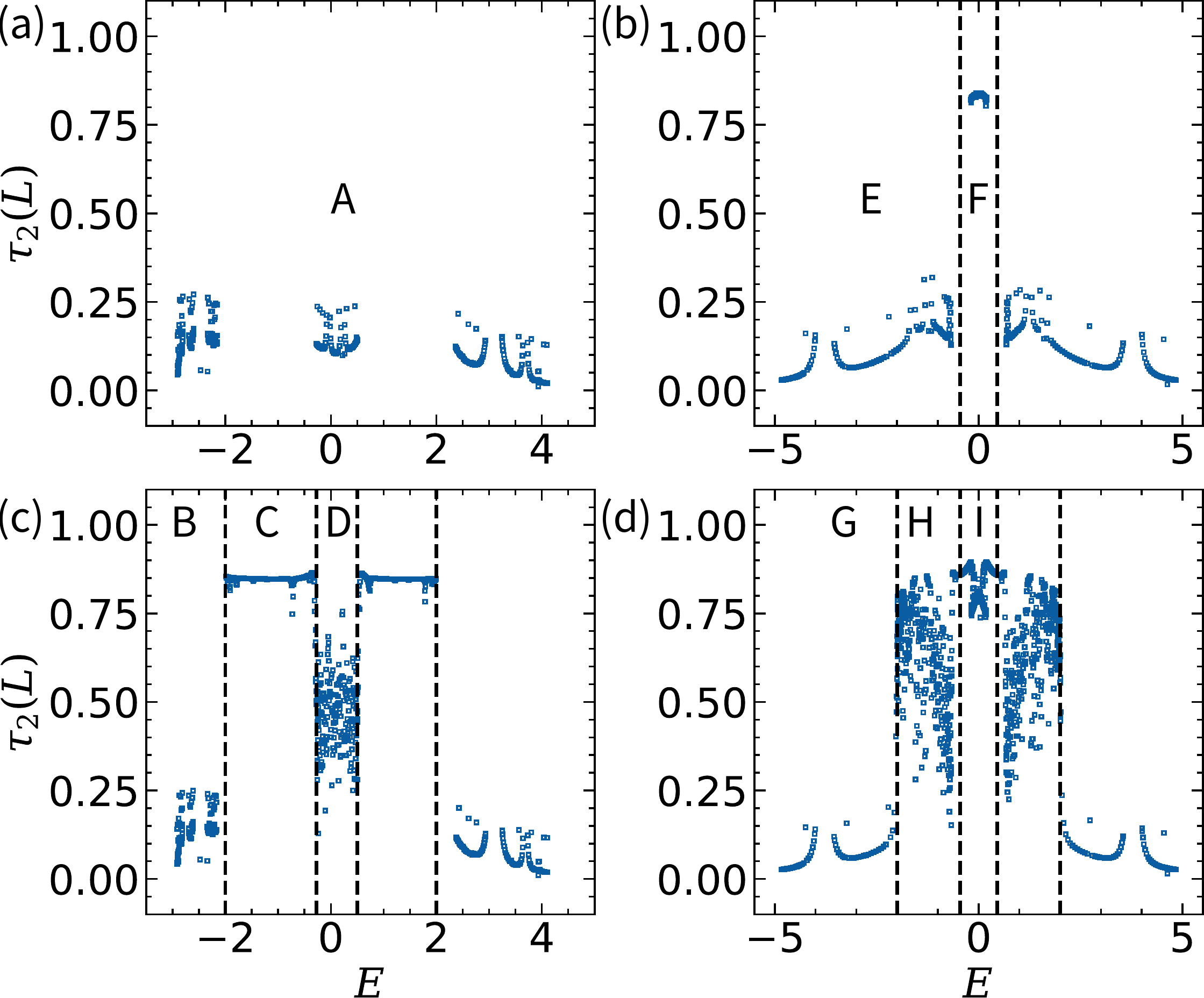}
\caption{
The FD $\tau_2(L)$ versus energy $E$ for model (a) general AAH model and (b) quasiperiodic mosaic model; 
(c) coupled General AAH model and (d) coupled quasiperiodic mosaic model. 
We use $L = 2F_{15} =  1220$, $t_\text{v} = 0.1$, $V=1.5$ for (a)(c) and $V=1.1$ for (b)(d). 
The dashed lines in panel (b) correspond to energy $E=\pm 0.5$; in panel (c) correspond to $E=\pm 2$, $E=-0.27$, and $E=0.5$; in panel (d) correspond to $E=\pm 2$ and $E=\pm 0.5$.
The CP is in the regimes D and H with $\tau \sim 0.25 - 0.75$
}
\label{fig-potential-detail}
\end{figure}

\begin{figure}[!htbp]
\includegraphics[width=0.48\textwidth]{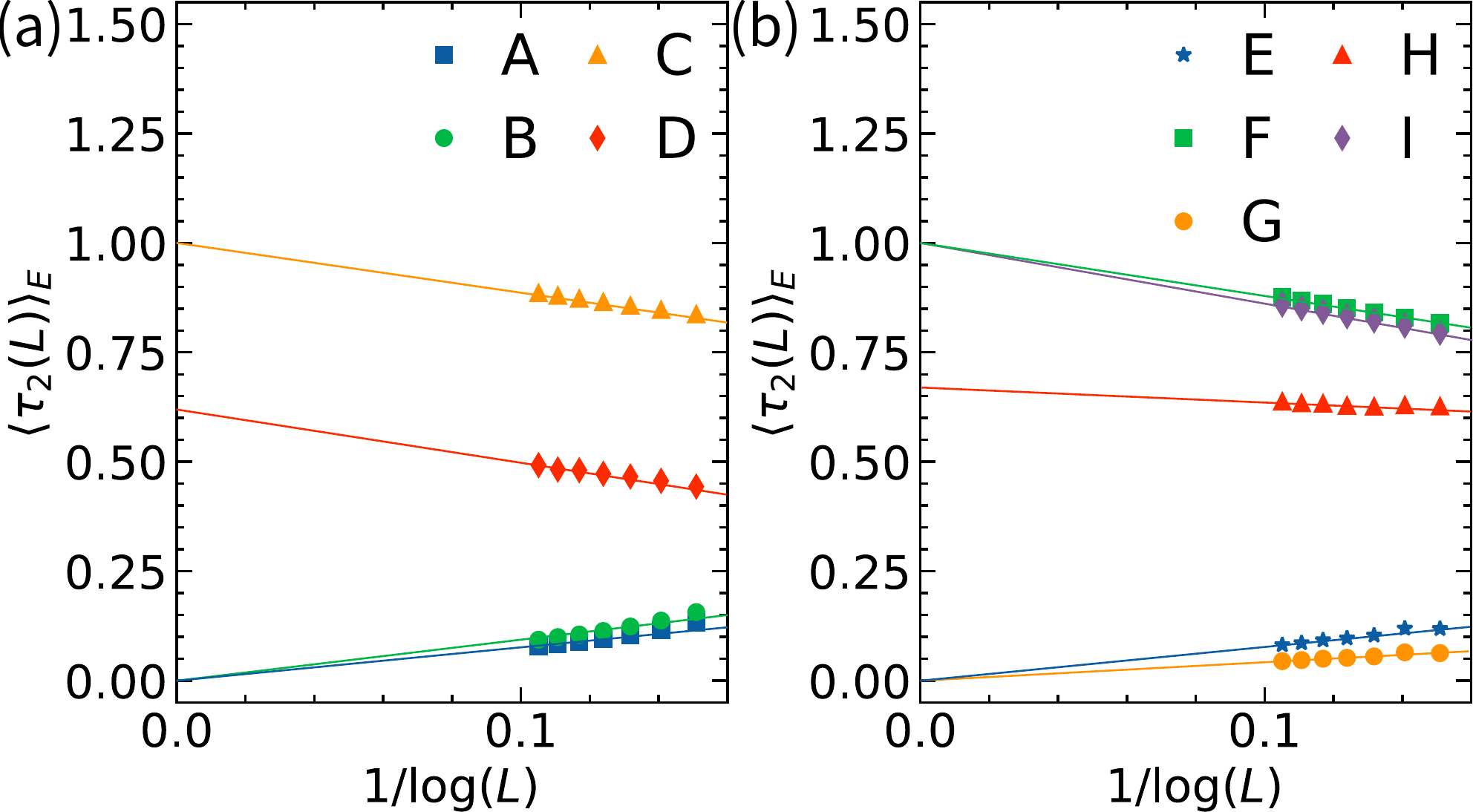}
\caption{
The mean value of $\langle \tau_2(L)\rangle_E$ in different energy regimes that corresponds to Fig. \ref{fig-potential-detail} against the inverse of $\log(L)$ with $L=2F_n$ the number of lattice sites.
$\langle \tau_2(L)\rangle_E$ in energy regimes A, B, E, and G approaching zero.
$\langle \tau_2(L)\rangle_E$ in energy regimes C, F, and I approaching unity. $\langle \tau_2(L)\rangle_E$ in energy regimes D and H approaching value between zero and unity, signaling the CP. }
\label{fig-potential-scale}
\end{figure}

A more precise plot of FD versus energy $E$ for this model is presented in Fig. \ref{fig-potential-detail}. We also presented the finite-size scaling of the FD in Fig. \ref{fig-potential-scale}.
For the GAAH chain and coupled GAAH chain, we use $V=1.5$. In this case, the states in the GAAH model are all localized with $\tau_2(L)<0.25$ and $\lim_{L\rightarrow\infty}\tau_2(L) = 0$ (see Fig. \ref{fig-potential-detail} (a) and Fig. \ref{fig-potential-scale} (a)).
For coupled GAAH model, the states in regime B are localized ( $\tau_2(L)<0.25$ and $\lim_{L\rightarrow\infty}\tau_2(L) = 0$); the states in regime C are extended ( $\tau_2(L)>0.75$ and $\lim_{L\rightarrow\infty}\tau_2(L) = 1$); the states in regime D are critical ( $0.25<\tau_2(L)<0.75$ and $0<\lim_{L\rightarrow\infty}\tau_2(L) <1$).
It should be noted that regime D is the overlapped spectra between the GAAH chain and purely extended chains.
The results for the Mosaic AAH chain and coupled Mosaic AAH chains exhibit the same features.
All those results suggest the approach we proposed to the CPs is general and should be applicable to the much wider family of coupled quasiperiodic chains.

\subsection{Inter-chain duality in coupled quasiperiodic chains}\label{sec-dual-aah}
As we have verified that this approach is general and is insensitive to the form of quasiperiodic potential and local inter-chain couple, it provides a general recipe to realize different types of CP on demand in the future.
It is possible to construct a physical model with self-dual symmetry by two coupled chains, where all states are critical which is similar to that at the critical point of the AAH model. 
The Hamiltonian reads as
\begin{equation}
H = H_{\mathrm{b}} + H_{\mathrm{a}} + H_\text{c},
\label{eq-HamDcc}
\end{equation}
where 
\begin{equation}
H_{\mathrm{b}} = \sum_m (b_m^{\dagger}b_{m+1} + \text{h.c.})  + 2V\cos(2\pi\beta m) b_m^{\dagger}b_{m},
\end{equation}
and 
\begin{equation}
H_{\mathrm{a}} = \sum_m J (a_m^{\dagger}a_{m+1} + \text{h.c.}) + 2\cos(2\pi\beta m)a_m^{\dagger}a_{m}.
\label{eq-Ham-dual}
\end{equation} 
Each chain, independently, corresponds to the physics in an AAH model, as discussed before. The inter-chain coupling $H_c$ is set to be the same as that in Eq. \ref{eq-minimal-model}.

\begin{figure}[!htbp]
\includegraphics[width=0.48\textwidth]{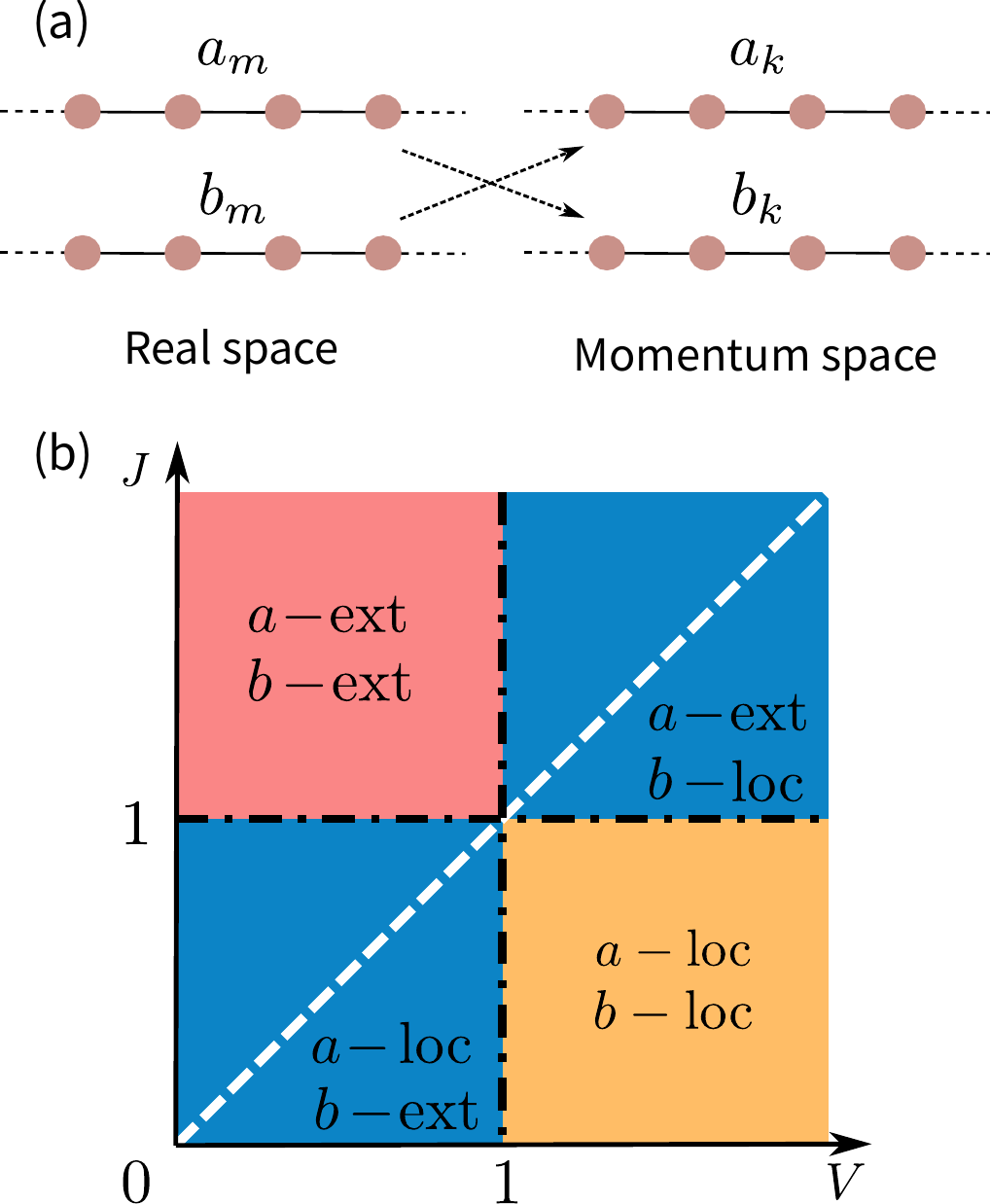}
\caption{
(a) Self-dual mapping between the two coupled chains with quasiperiodic potential.
(b) The phase diagram of the model in Eq. \ref{eq-Ham-dual} without inter-chain coupling,  $t_\text{v}=0$. Here $a$($b$)-ext and $a$($b$)-loc are abbreviate for extended and localized states in $a$ ($b$) modes.
The self-dual line between the two chains is given by $J=V$ and the two vertical dashed lines are the self-dual points in each chain, with $J=1$ and $V=1$.
In the presence of inter-chain coupling $H_c$, the overlapped spectra in the two diagonal blocks may yield CP.
Especially, all states in the dual line $J=V$ are critical.}
\label{fig-schematic-dual}
\end{figure}
Without inter-chain coupling $H_c$, this model is reduced to two independent AAH models. 
Thus the states in Eq. \ref{eq-HamDcc} are either localized or extended, depending on the parameters $J$ and $V$, with critical points at $J_c =1$ and $V_c =1$. 
However, this model also has an intriguing symmetry, dubbed as inter-chain dual symmetry, in which the Fourier transformation of $a_m$ and $b_m$ in real space is changed to $b_k$ and $a_k$ in the momentum space.
This symmetry is schematically shown in Fig. \ref{fig-schematic-dual} (a).
We may even regard the $b$ mode as the Fourier component of the $a$ mode in the Hamiltonian. In this way, we find a self-dual symmetry between the two chains at $J=V$, which forms a self-dual line in the parameter space, instead of a single self-dual point in the AAH model. 
Thus the total phase diagram is shown in Fig. \ref{fig-schematic-dual}, in which the vertical and horizon lines represent the self-dual in each AAH model and the diagonal line $J=V$ represents the inter-chain self-dual condition. We have four different phases. 
When $J <1$, $V <1$ or $J > 1$, $V <1$ for the two diagonal blocks, states in one of the chains are localized and in the other chains are extended, leading to possible coexistence of localized and extended phases in the energy space. 
Especially, at the self-dual line $J=V$, the two chains should have the same equation of motion, thus the spectra are exactly overlapped. 
Otherwise, their spectra are partially overlapped. 
In the off-diagonal blocks with $J > 1$ and $V< 1$, all states are extended; and in regime $J < 1$ and $V > 1$, all states are localized, which are trivial even in the presence of inter-chain coupling.
Thus, from the picture discussed in this work, we expect the presence of inter-chain coupling can change the coexisted localized and extended states into critical states in the two diagonal blocks. 

\begin{figure}[!htbp]
\includegraphics[width=0.48\textwidth]{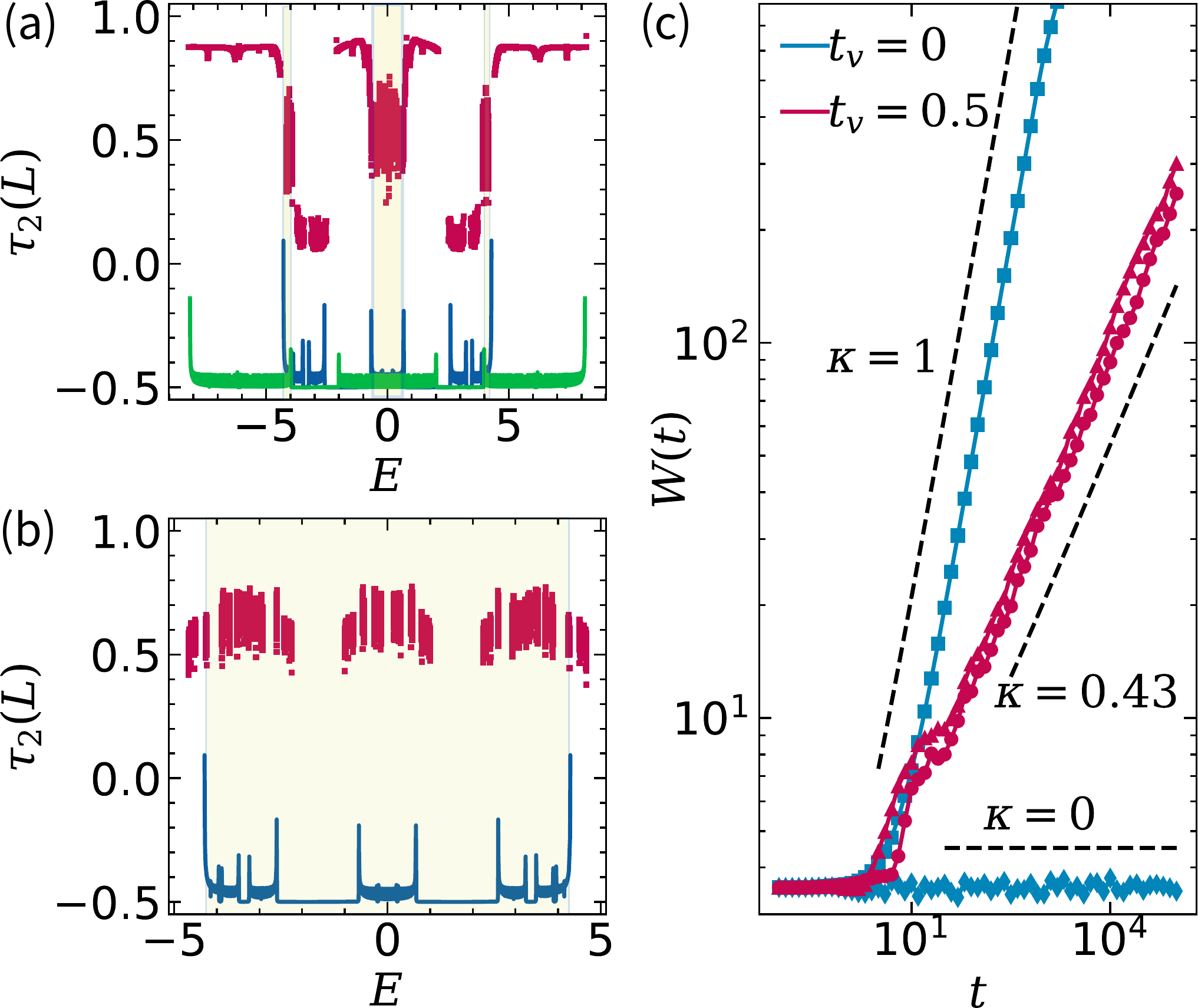}
\caption{(a) and (b) FD $\tau_2(L)$ for the dual coupled chains model (\ref{eq-Ham-dual}) with $L=2F_{19} = 8362$, $t_\text{v}=0.5$, $V=2$. Here we use red squares for $\tau_2(L)$ and the blue or green lines for the density of states in the AAH model.
The overlapped spectra are shown by the yellow shadows.
The free chain hopping strength is (a) $J=4$; (b) $J=2$, thus in (b) the system has self-dual symmetry, and all states are critical.  
(c) Evolution of mean-square $W(t)$ with different initial states. 
The square dotted lines for $\gamma = \mathrm{ext}$ and the triangle dotted lines for $\gamma = \mathrm{loc}$. The black dashed lines with different slopes serve as a guide for the eyes. 
We use $L = 2F_{19} = 8362$, $J = V =2$ in this panel.
}       
\label{fig-dual}
\end{figure}

It is interesting to find that in the dual line, $H_c$ could also be self-dual \footnote{Other forms of inter-chain coupling without self-duality can also induce CP (see the discussion in Sec. \ref{sec-general}.)}, which can be seen from the following Fourier transformation 
\begin{equation}
\sum_{i} a_i^\dagger b_i + \text{h.c.} \rightarrow \sum_{k} 
a_k^\dagger b_k + \text{h.c.}. 
\label{eq-fouriertv}
\end{equation}
Thus we expect that all states in this line are critical in the presence of inter-chain coupling.
This can be seen from the special point at $J=V=1$, at which it is well-known that all states without $H_c$ are critical. 
Otherwise, when $J\ne V$, only the overlapped spectra can turn into critical. To examine the criticality of these states, we calculate the FD $\tau_2(L)$ for $V=2$ and $J=4$, $J=2$ in Fig. \ref{fig-dual} (a) and (b), respectively. 
At the self-dual condition, we find all $\tau_2(L)$ are in the range of $(0.4, 0.8)$ for criticality. 

Furthermore, we consider the dynamics of the following initial Gaussian packet ($\gamma =$ loc, ext) 
\begin{equation}
    |\psi^{\gamma}(0)\rangle \propto \sum_m \mathrm{e}^{-\frac{(m-m_0)^2}{(2\sigma^2)}} (\delta_{\gamma,\mathrm{loc}}b_m^{\dagger} + \delta_{\gamma,\mathrm{ext}}a_m^{\dagger})|0\rangle,
\label{eq-wave-packet-initial}
\end{equation}
with wave packet width $\sigma=5$. These dynamics can be performed in experiments \citep{roati_anderson_2008,luschen_spme_2018}. 
The mean-square of the packet $W(t) =\sqrt{ \langle x^2\rangle - \langle x\rangle^2}$ with $x = \sum_m m(a_m^{\dagger}a_m + b_m^{\dagger}b_m)$ in the long-time limit yields
 \begin{equation}
W(t) = \left(\sum_{m,\mu} |m-\langle x\rangle|^2 |\psi_{m \mu}|^2\right)^{1/2} \sim t^{2\kappa}.
\end{equation}
In general, the dynamical exponent $\kappa = 0$, $\kappa = 1$ and $0 < \kappa < 1$ correspond to localized states, extended states, and critical states, respectively \cite{Zhang2012Hyperdiffusion, Abe1987Fractal, hiramoto_dynamics_1988,wang_realization_2020}. 
The dynamics of this wave packet are presented in Fig. \ref{fig-dual} (c). 
We find that when the system has self-dual symmetry along the dual line, $\kappa \sim 0.43$, which is comparable with the value in the AAH model with $\kappa = 1/2$ \citep{hiramoto_dynamics_1988,geisel_new_1991}, which is a unique signature of criticality \citep{geisel_new_1991, wang_one-dimensional_2020, Tong2002Electronic}.
However, in the fully localized state, $\kappa = 0$, and in the condition when part of spectra are extended, $\kappa=1$.
We also discuss the dynamical exponent with some other self-dual parameters $J=V$, which are presented in Appendix \ref{sec-appendix-wavepacket}, showing that $\kappa \sim 0.5$ depends weakly on these parameters. 

\subsection{Several remarks}
\label{sec-remarks}

This idea may also be generalized to the other coupled quasiperiodic chains with some types of self-duality. 
Several remarks about these possibilities are in order.
Firstly, it may also be used to study the possible duality in multiple coupled chains, in which different types of inter-chain duality can yield much more complicated CPs in the parameter space. 
For example, for three coupled chains, the duality between $H_1$ and $H_2$, $H_1$ and $H_3$, and $H_2$ and $H_3$ can yield three different inter-chain dualities, giving rise to a much richer phase diagram. 
Secondly, it can be used to study the possible CP in the optical ladder lattice with gauge potential, which has recently been realized in experiments and has been a subject of theoretical research \cite{Atala2014Meissner, Livi2016Synthetic, Qiao2021Quantum, Orignac2001Meissner, Piraud2015Vortex}. 
This ladder structure is essential to the two coupled chains presented in Fig. \ref{fig-approach} (a) with flux $\phi$ per plaquette, and the Hamiltonian can be written as \cite{Atala2014Meissner, Livi2016Synthetic}
\begin{eqnarray}
H = && \sum_{m} -J(a_{m,1}^\dagger a_{m+1,1} + a_{m,2}^\dagger a_{m+1,2})  \nonumber \\
  && + t_\text{v} a_{m, 1}^\dagger a_{m, 2} e^{i m\phi} + \text{h.c.}.
\end{eqnarray}
This model may undergo a transition from the Meissner phase and Abrikosov vortex phase, which can be directly manifested from the chiral current \cite{Atala2014Meissner}, and the critical coupling strength is given by $(t_\text{v}/J)_c = 2J\tan(\phi/2) \sin(\phi/2)$. 
For $|t_\text{v}| \ll |J|$, it is a Meissner phase; however, when $|t_v| \gg |J|$ is dominated, it is a vortex phase. 
These physics are analogies to the physics in type-II superconductors, in which vortex states exist when the magnetic field exceeds some critical field $H_{c_1}$.
This model can be written in a different form, assuming  $a_{m,1} \rightarrow a_{m,1} e^{i m\phi/2}$ and $a_{m,2} \rightarrow a_{m,2} e^{-im\phi/2}$, then we obtain 
\begin{eqnarray}
H = && \sum_{m} -(J a_{m,1}^\dagger a_{m+1,1}e^{-i\phi/2} + a_{m,2}^\dagger a_{j+1,2}e^{i\phi/2})  \nonumber \\
  && + t_\text{v} a_{m,1}^\dagger a_{m,2} + \text{h.c.}.
\end{eqnarray}
In a single chain, the phase in the hopping can be gauged out, thus is not important. 
However, in ladder geometry, it is essential for new behaviors.
In Ref. \cite{Qiao2021Quantum}, the authors have investigated the quantum phase transition and Bose-Einstein condensation in this ladder model in the presence of on-site many-body interaction, which led to a rich phase diagram. 
Furthermore, the hopping term can be mapped to the $\cos(2\pi\beta k + \phi/2) a_{k,1}^\dagger a_{k,1}$ and $\cos(2\pi\beta k - \phi/2) a_{k,2}^\dagger a_{k,2}$ in the momentum space (see Eq. \ref{eq-aah} and Eq. \ref{eq-aah-momentum}), while the coupling term is invariant (see Eq. \ref{eq-fouriertv}). 
Thus, if we included a quasiperiodic potential in the above model, it can be self-dual having inter-chain or intra-chain duality.
Therefore, this model may provide fertile ground for exploring the interplay between AL, criticality, Meissner-vortex transition, and many-body effects \cite{Qiao2021Quantum, Orignac2001Meissner, Piraud2015Vortex}, which should support various novel phase transitions. 

\section{
The applicability of perturbation theory and the effective unbounded potential}\label{sec-mechansim}
The physics for the emergence of CP in the overlapped spectra in the coupled quasiperiodic chains is novel and has great generality. 
However, it is still unclear to us what caused the physics in the overlapped and un-overlapped spectrum to be so different. 
Similar incertitude also exists in the previous literature for CP, which hinders the realization of CP in a more general way. 
In this section, we mainly focus on the minimal model for simplification, while all those results and arguments can also be applied to other coupled quasiperiodic chains.

\subsection{Fidelity between decoupled chains and coupled chains in the minimal models}\label{sec-fidelity}
\begin{figure}[!htbp]
\centering
\includegraphics[width=0.48\textwidth]{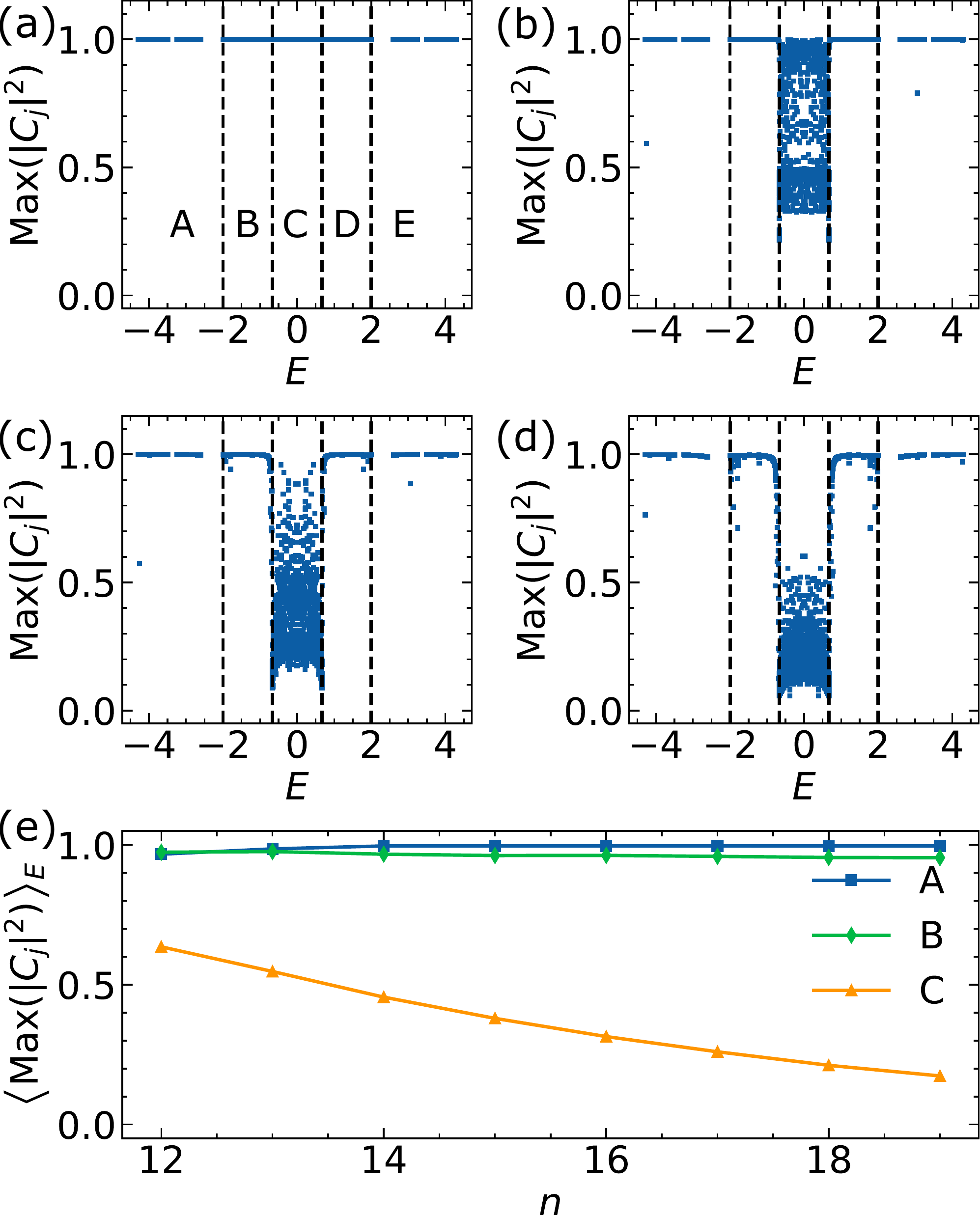}
\caption{ (a)-(d) $\mathrm{Max}(|C_j|^2)$ for all states with $L = 2F_{18} = 5168$ and $V = 2$. the inter-chain coupling is chosen to be (a) $t_\text{v}=0$, (b) $t_\text{v}=0.01$, (c) $t_\text{v}=0.05$, (d) $t_\text{v}=0.1$.
In the un-overlapped regimes, the hybridization between extended states and localized states is not significant, yielding $\mathrm{Max}(|C_j|^2) 
\sim 1$. In the overlapped regime, the coupling between these different states yields $\mathrm{Max}(|C_j|^2) \ll 1$. 
The black dashed lines denote the energy $|E| = \pm 0.67$, $|E|=\pm 2$.
(e) Averaged maximum $\langle |C_j|^2 \rangle_E$ in specific energy intervals A, B, and C versus Fibonacci index $n$ with $V=2$, $t_\text{v}=0.1$.
In the CP, $\mathrm{Max}(|C_j|^2)$ decreases to zero with the increase of the Fibonacci index.
In the un-overlapped spectra, this value is almost independent of system size.
}
\label{fig-fidelity}
\end{figure}

In order to have an insight into this question, it is crucial to have some knowledge of the relationship between the states $|\phi_{j'}^{0}\rangle$ in the decoupled chains ($t_\text{v} = 0$), and the states $|\phi_j^{t_\text{v}}\rangle$ in the coupled chains ($t_\text{v} \neq 0$). 
We use the overlap between these states to determine the influence of $H_c$, which is defined as 
\begin{equation}
C_{j'j} = \langle \phi_{j'}^{ 0}\mid\phi_j^{t_\text{v}} \rangle,
\end{equation}
between the two states indexed by $j$
and $j'$. Here $C$ forms an orthogonal matrix, in regarding that all eigenvalues can be represented using real vectors, thus $C^T C = \mathbb{I}$.
This matrix directly reflects the reconstruction of the eigenvalues in terms of inter-chain coupling $H_c$. For convenience, we only examine 
\begin{equation}
    \mathrm{Max}(|C_{j}|^2) = \max\{|C_{j'j}|^2|,\ j' = 1, 2, \cdots\}
\end{equation}
to characterize the similarity between states in these two chains for a given $j$.
If the state $|\phi_j^{t_\text{v}}\rangle$ is similar to a state in the decoupled chains, then $\mathrm{Max}(|C_{j}|^2) \sim 1$. Otherwise, $\mathrm{Max}(|C_{j}|^2)$ will less than unity. 
Our results are summarized in Fig. \ref{fig-fidelity}, where we find $\mathrm{Max}(|C_{j}|^2)$ is approximately equal to unity for the states in energy regimes A, B, D, E. 
Nevertheless, $\mathrm{Max}(|C_{j}|^2)$ of the states in energy regime C is less than unity and has a large fluctuation.
Especially, $\mathrm{Max}(|C_{j}|^2)$ in regime C decrease to zero with the increasing of inter-chain coupling (see Fig. \ref{fig-fidelity} (a) to (d)). 
We further examine the dependence of $\mathrm{Max}(|C_{j}|^2)$ on the size of the system, which is presented in Fig. \ref{fig-fidelity} (e). 
We find that $\left\langle \mathrm{Max}(|C_{j}|^2)\right\rangle_E$ in regime C decreases to zero with the increasing of the system size $L =2F_n$, which in A and B (and D and E), they are unchanged. 
These features suggest that the wave functions in the overlapped spectra are completely reconstructed, while the wave functions in the un-overlapped spectra are similar to that in the decoupled chains.
Therefore, a perturbative argument should be applied to the states in the un-overlapped spectra, which will be discussed below.

\subsection{The applicability of perturbation theory}
Here, we propose an argument in order to explain the physics in Sec. \ref{sec-fidelity}. In Fig. \ref{fig-fidelity}, the states in the un-overlapped spectra of the coupled chains are similar to the states in the decoupled chains.
This is related to the applicability of perturbation theory, in which the localized and extended states can only be weakly coupled when their energies are separated by a finite gap $\Delta$.
Let the extended and localized states of the decoupled chains as $|\phi_{j,\text{e}}^{0}\rangle$ and $|\phi_{i,\text{l}}^{0}\rangle$ ($i$ and $j$ are the labels for eigenvalue here), which have an amplitude of $1/\sqrt{L}$ and $1/\sqrt{W}$, respectively, with $W$ being their width, then 
\begin{equation} |\phi_{j,\text{e}}^{t_\text{v}} \rangle = |\phi_{j,\text{e}}^{0} \rangle +  \sum_{i} \frac{\langle \phi_{i,\text{l}}^{0} |H_\text{c}| \phi_{j,\text{e}}^{0}\rangle}{E_{j,\text{e}}^{0} - E_{i,\text{l}}^{0}} |\phi_{i,\text{l}}^{0} \rangle  + \dots,
    \label{eq-perturbation-extend}
\end{equation}
and 
\begin{equation}
    |\phi_{j,\text{l}}^{t_\text{v}} \rangle = |\phi_{j,\text{l}}^{0} \rangle +  \sum_{i} \frac{\langle \phi_{i,\text{e}}^{0}  |H_\text{c}| \phi_{j,\text{l}}^{0}  \rangle}{E_{j,\text{l}}^{0}  - E_{i,\text{e}}^{0} }|\phi_{i,\text{e}}^{0}  \rangle  + \dots,
    \label{eq-perturbation-localize}
\end{equation}
with $|\phi_{j,\text{e}}^{t_\text{v}}  \rangle$ and $|\phi_{j,\text{l}}^{t_\text{v}} \rangle$ are the wave functions for the coupled chains.
We find that the coupling matrix elements (in $d=1$)
\begin{equation}
|\frac{\langle \phi_{i,\text{e}}^{0} |H_\text{c}| \phi_{i,\text{l}}^{0}\rangle}{E_{j,\text{l}}^{0} - E_{i,\text{e}}^{0} }| \leq \frac{t_\text{v}}{\Delta} \cdot \sqrt{\frac{W}{L}}.
\label{eq-WLscaling}
\end{equation}
This inequality can be understand by assuming $\phi_{\text{e}} = \sqrt{2/L}\sin(j \pi x/ L)$ and $\phi_{\text{l}} = (2/\pi W^2)^{1/4} \exp(-(x-L/2)^2/W^2)$, which are normalized. 
If $W \ll L$, the direct overlap between them is given by 
\begin{equation}
    \int \phi_{\text{e}}^* \phi_{\text{l}} dx = (8\pi)^{1/4} \sqrt{\frac{W}{L}} \exp(-\frac{\pi^2 W^2}{4L^2}) \sim \sqrt{\frac{W}{L}},
\end{equation}
which directly yields the scaling in Eq. \ref{eq-WLscaling}.
The contribution of the perturbation to the wave function can be estimated via (using $\sum_j 1/L = 1$)
\begin{equation}
    \left(\sum_i \frac{\langle \phi_{i,\text{e}}^{0} |H_\text{c}| \phi_{i,\text{l}}^{0}\rangle}{E_{j,\text{l}}^{0} - E_{i,\text{e}}^{0} }\right)^2 \leq 
    \sum_{j} (\frac{t_\text{v}}{\Delta})^2 \frac{W}{L} = \frac{t_\text{v}^2 W}{\Delta^2} \ll 1,
\end{equation}
as long as $t_\text{v} \sqrt{W}/\Delta$ is sufficiently small, which can be realized with a weak inter-chain coupling $t_\text{v}$ (e.g. $t_\text{v} = 0.1$) or a large detuning $\Delta$.
The higher-order contribution of the perturbation series on the wave function $|\phi^{t_\text{v}}_{j,\text{l}} \rangle$ and $|\phi^{t_\text{v}}_{j,\text{e}} \rangle$ would be much smaller using $|\frac{\langle \phi_{i,\text{e}}^{0} |H_\text{c}| \phi_{i,\text{l}}^{0}\rangle}{E_{j,\text{l}}^{0} - E_{i,\text{e}}^{0} }| \leq \frac{t_\text{v}}{\Delta} \cdot (\sqrt{\frac{W}{L}})$.

Thus, the new wave function in the presence of coupling is still approximately given by $|\phi_{j,\text{l}}^0\rangle$ or $|\phi_{j,\text{l}}^0\rangle$.
This argument explains why the states in the un-overlapped spectra are so similar to the states in the decoupled chains, thus
\begin{equation}
    |\langle \phi_{i,\text{e}}^{t_\text{v}}|\phi_{i,\text{e}}^0\rangle|,\quad |\langle \phi_{i,\text{l}}^{t_\text{v}}|\phi_{i,\text{l}}^0\rangle|  \sim 1.
    \label{eq-fidelity}
\end{equation}
And it also explains why all the states maintain localized or extended in the un-overlapped spectra regime (e.g. $\tau_2(L)$ is unchanged when $|E| > 2$ in Fig. \ref{fig-minimal} (b) - (d)). Namely, localized states are still localized and extended states are still extended in the un-overlapped spectra in the presence of inter-chain coupling.
We justify this prediction in Fig. \ref{fig-fidelity} (e). 

However, there is no chance that this argument can be applied to the physics in the overlapped spectra, where the denominator of the series may approach zero, yielding divergent coefficients in Eq. \ref{eq-perturbation-extend} and Eq. \ref{eq-perturbation-localize}.
Thus one needs to consider all the higher-order terms, which make the wave functions completely different from that in the decoupled chains. It is expected 
\begin{equation}
\begin{aligned}
&|\langle \phi_{j,\text{e}}^{t_\text{v}}|\phi_{i,\text{e}}^{0}\rangle|,\quad |\langle \phi_{j,\text{e}}^{t_\text{v}}|\phi_{i,\text{l}}^{0}\rangle| \ll 1 ,\\
&|\langle \phi_{j,\text{l}}^{t_\text{v}}|\phi_{i,\text{e}}^{0}\rangle|,\quad |\langle \phi_{j,\text{l}}^{t_\text{v}}|\phi_{i,\text{l}}^{0}\rangle|  \ll 1,
\end{aligned}
\end{equation}
for all states in the overlapped spectra regime, which should decrease to zero with the increasing of system length $L$.
Namely, all states are reconstructed with this weak inter-chain coupling for the CP.
This argument is a resemblance to that in the original Anderson transition \citep{anderson_absence_1958}, in which the divergence of the higher-order series signals the phase transition from the extended states to the localized states.
We justify these results in Fig. \ref{fig-fidelity} (e). 

\subsection{The unbounded potential}

Although, the perturbation argument successfully explains the physics in the un-overlapped spectra.
There are still two questions that need to be solved. The first one is why the reconstruction of wave functions in the overlapped spectra should generate multifractal wave functions.
The second one is that could we generate CP from coupled chains where the quasiperiodic potentials are replaced by random potentials.  
The key point to answer these two questions is the emergence of effective unbounded potential, in which quasiperiodic potential and disordered potential will exhibit totally different behaviors.
To clarify these issues much more clearly, we rewrite the Schr\"{o}dinger equation of the Hamiltonian in Eq. \ref{eq-H-general} in the following way as 
\begin{equation}
\begin{aligned}
H_1 |\Psi_1\rangle  + H_c |\Psi_2\rangle = E |\Psi_1\rangle \\
H_c^{\dagger} |\Psi_1\rangle  + H_2 |\Psi_2\rangle = E |\Psi_2\rangle,
\end{aligned}
\end{equation} 
This equation has a solution as 
\begin{equation}
(H_i + \Sigma_i(E)) |\Psi_i\rangle = E |\Psi_i\rangle, \quad i=1, 2, 
\label{eq-project}
 \end{equation}
where the self-energy or effective potential can be written as 
\begin{equation}
\Sigma_1 = H_c\frac{1}{E-H_2}H_c^{\dagger}, \quad 
\Sigma_2 = H_c^{\dagger}\frac{1}{E-H_1} H_c.
\label{eq-selfenergy}
\end{equation}
The difference between states in the coupled chains with uncorrelated random potential or quasiperiodic potential can be seen later.
For simplification, we set states in $H_2$ to be fully extended, such as $H_2 = \sum_m (a_{m+1}^{\dagger}a_{m} + \mathrm{h.c.})$ and states in $H_1$ to be fully localized, such as $ H_1 = \sum_m V_m b_m^{\dagger}b_m$ with $V_m = V\cos(2\pi\alpha m)$ quasiperiodic or $V_m$ a random number taken from a uniform distribution $U(-V/2,V/2)$. 
We further assume the inter-chain coupling to be local, such as $H_c = \sum_m t_\text{v}(a_m^{\dagger}b_m + \mathrm{h.c.})$.
Then the eigenstates and eigenvalues of $H_1$ are $E_i^{(1)} = V_i$ and $|i^{(1)}\rangle = b^{\dagger}_i| 0\rangle$; and the eigenstates and eigenvalues of $H_2$ are $ E_k^{(2)} = 2\cos(k) $ and $|k^{(2)}\rangle = \sum_m \frac{e^{i k m}}{\sqrt{L}} a^{\dagger}_m| 0\rangle$ with $k=2 \pi n/L$, with $n$ being an integer in $[1,L]$.
We insert the eigenstates and eigenvalue into Eq. \ref{eq-project}, then
\begin{eqnarray}
(H_1 + t_\text{v}^2 \sum_{m,n} G(m,n;E) a_m^{\dagger}a_n)|\Psi_1\rangle  &= E|\Psi_1\rangle, \label{eq-project-a}\\
(H_2 + t_\text{v}^2 \sum_m \frac{1}{E-V_m}b^{\dagger}_m b_m )|\Psi_2\rangle &= E|\Psi_2\rangle,
\label{eq-project-b} 
\end{eqnarray}
with $G(m,n;E) =  \sum_k \frac{\langle m| k^{(2)}\rangle\langle k^{(2)}| n \rangle}{E-E_k^{(2)}} $ the Green function of $H_2$. Its expression in the large $L$ limit is 
\begin{equation}
    G(m,n;E) = \frac{1}{\sqrt{E^2-4}} \left(\sqrt{\frac{E^2}{4}-1} - \frac{E}{2}\right)^{|m-n|}.
    \label{eq-green-function}
\end{equation}
This result is well-known and can be found in the textbook \cite{Hoffmanbook}.
When $|E| < 2$, we can find $|\sqrt{\frac{E^2}{4}-1} - \frac{E}{2}| =1 $ and $G(m,n;E)$ does not decay with the distance; yet when $|E| > 2$, $G(m,n;E)$ should be exponentially localized since $|\sqrt{\frac{E^2}{4}-1} - \frac{E}{2}| \sim 1/|E| < 1$. 
With all those results, we can further explain the fate of these states from the perspective of overlapped spectra and un-overlapped spectra, which can host different states (see Fig. \ref{fig-approach}). 

Firstly, we consider the results in the un-overlapped spectra regime. 
In this case, we only need to consider Eq. \ref{eq-project-a} or Eq. \ref{eq-project-b} respectively.
For $H_1$ in Eq. \ref{eq-project-a}, the Green's function 
$G(m,n; E)$ is always short-range correlated, thus no matter which kind of potentials have been used in $H_1$, states in this equation should always be localized.
In contrast, in Eq. \ref{eq-project-b}, we may approximate $1/(E - V_m) \simeq 1/E - V_m/E^2$, thus when $V_m$ is a random potential, states in $H_2$ are localized; 
however, when $V_m$ is the quasiperiodic potential, states in $H_2$ are extended when $t_\text{v}$ is weak enough, in the case that $H_2$ dominated. 
We may conclude that the inter-chain coupling could not effectively hybridize the localized states and extended states in the un-overlapped spectra for coupled chains with quasiperiodic potential, leaving them either localized or extended.
Whereas the inter-chain coupling localized the extended states by inducing small random disorders from the localized states for coupled chains with random potential, leaving all the states localized in the un-overlapped spectra.

Secondly, we consider the fate of states in the overlapped spectra regime. Here, we need to consider both Eq. \ref{eq-project-a} and Eq. \ref{eq-project-b}.
In Eq. \ref{eq-project-a}, $G(m, n; E)$ is long-range correlated, leading to all-to-all hopping in the $H_1$ chain. 
Since these hoppings are all-to-all, they should delocalize the wave function of $|\Psi_1\rangle$.
On the other hand, in Eq. \ref{eq-project-b}, the denominator of $1/(E - V_m)$ may approach zero at some sites, leading to unbounded potential, in which the wave function of $|\Psi_2\rangle$ can not be extended.
With these two competition effects, it is hard to conclude the fate of these phases in the presence of inter-chain coupling. 
However, for the unbounded potential, it has been clearly shown \citep{simon_trace_1989, kirsch_one-dimensional_1990} that it rules out the possibility of continuous spectra for the extended states in one-dimensional systems, leaving the states to be either localized or critical, depending on the forms of $V_m$. 
Furthermore, recently, Liu \textit{et al.} have studied the unbounded potential $V(n) \propto 1/(1 - 
a \cos(2\pi\alpha n))$ (see Eq. \ref{eq-gaah}) with $a>1$, which is closely related to the potential in Eq. \ref{eq-project-b}, using the Avila's global theory \cite{Avila2015Global} and yield the CP in the energy windows $|E|<2$. 
Thus we may have reasons to believe that in the overlapped spectra, the reconstruction of wave functions through resonance coupling may lead to CP for models with quasiperiodic potentials. However, all the states for models with random potential should be localized according to the scaling argument \citep{abrahams_scaling_1979}.

Combining the arguments above, we explain the reason for the emergence of CP in coupled quasiperiodic chains and answer the question about the fate of states in coupled chains with random potentials. 
This analysis may also illustrate the fundamental difference between the quasiperiodic potential and random potential. 
It is necessary to emphasize that this conclusion does not contradict the long-range correlated disorder result because the effective potential $\Sigma_i \sim 1/(E-V_m)$ is also considered long-range correlated \cite{Moura1998Delocalization} supporting AL in 1D.  
We also emphasize that while our argument in this section is based on the minimal model in Sec. \ref{sec-minimalmodel}, it can be generalized to other coupled quasiperiodic chains directly.

\section{The possible CP in bichromatic optical lattice}\label{sec-observation-realization}

\begin{figure}[!htbp]
\includegraphics[width=0.48\textwidth]{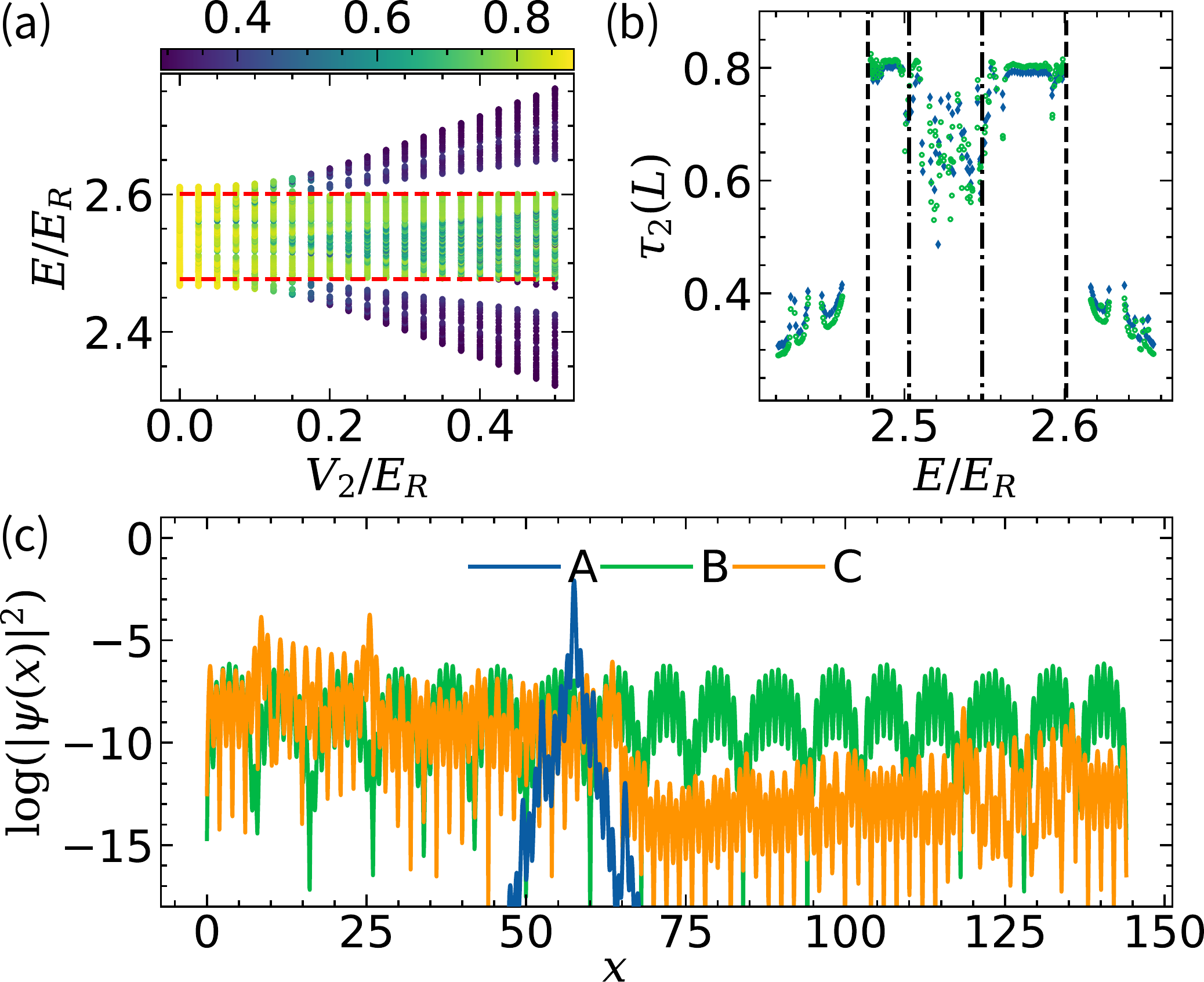}
\caption{
(a) The FD $\tau_2(L)$ for the states in the s-band of the continuous model in Eq. \ref{eq-continous-model} with $L= 144 a$, $\Omega/E_R=0.01$, $V_1/E_R=8$ for different incommensurate lattice strength $V_2$. The red dashed lines are the s-band edges at $V_2/E_R = 0$. 
(b) FD $\tau_2(L)$ for $V_2/E_R=0.25$ with other parameters the same as panel (a). The blue diamonds (green squares) correspond to $L=89a$ ($L=144a$). The dashed and dashed-dotted lines label the different phases; see Fig. \ref{fig-coupling}. 
(c) The logarithm of wave function amplitude $|\psi(x)|^2 = |\psi_{\uparrow}(x)|^2 + |\psi_{\downarrow}(x)|^2$ against $x$. Here eigenstates $\psi_{11}$ in A, $\psi_{71}$ in B and $\psi_{145}$ in C are chosen. }
\label{fig-continuous}
\end{figure}

While all these results are demonstrated based on tight-binding models, it is important to ask whether this CP can still be observed using the continuous Schr\"odinger equations, which may be readily realized using the quantum simulation platforms, such as ultracold atoms in optical lattice \citep{schreiber_observation_2015, Kohlert2019Observation}.
As pointed out in the previous literature \citep{Biddle2009Localization, li_mobility_2017}, the continuous model is not fully governed by the tight-binding description even when the strength of the optical lattice is very strong. 
Here we focus on a specific case that a two-level atom is placed on a spin-dependent bichromatic optical incommensurate lattice, in which the two hyperfine states are coupled by microwave or two-photon Raman coupling. 
We consider a one-dimensional model, assuming strong confinement along the other two transverse directions. 
Then the physics is described by the following continuous Hamiltonian
\begin{equation}
    H = -\frac{\hbar^2}{2m}\frac{\partial^2}{\partial x^2} + V_{\text{p}}(x) + 
    \begin{pmatrix}
    0  & \Omega/2\\
    \Omega/2 & V_{\text{in}}(x)
    \end{pmatrix},
    \label{eq-continous-model}
\end{equation}
with 
\begin{equation}
    V_\text{p}(x) = V_1\cos^2(k_1 x), \quad 
    V_{\text{in}}(x) = \frac{V_2}{2}\cos(2 k_2 x),
\end{equation}
with $k_1 = 2\pi/\lambda_\text{w}$, $\lambda_\text{w}$ is the wavelength of the light, $\beta = k_2/k_1$, $m$ is the atom mass, and $\hbar$ is the Planck constant divided by $2\pi$. $\Omega$ is the coupling between the two hyperfine states,
which plays the same role as $t_\text{v}$ in the coupled tight-binding models. 
The detailed realization of this model is presented in Appendix \ref{sec-appendix-experiment}. 
We set the recoil energy $E_R = \frac{\hbar^2 k_1^2}{2m}$ as the energy scale in the calculation and $a = \pi/k_1$ as the basic length scale. 

We solve the Schr\"{o}dinger equation of Eq. \ref{eq-continous-model} using finite difference method ($\delta x = 1/20$ and $L = 89$, $144$) in a finite system with open boundary condition. We find similar signatures that the CP emerges in the overlapped spectra, which is consistent with the prediction by the tight-binding model. 
The spectra of this model are presented in Fig. \ref{fig-continuous} (a) and the detailed plot of FD $\tau_2(L)$ with lattice size $L=89a$ and $L=144a$ are presented in Fig. \ref{fig-continuous} (b). 
These results are similar to that reported in Fig. \ref{fig-minimal} in the minimal model. 
In the overlapped spectra regime for $E \in (2.48, 2.60)$, we find $\tau_2(L) \sim 0.5 - 0.7$, while in the localized (extended ) states, we find $\tau_2(L)<0.4$ ($\tau_2(L)>0.75$). 
In Fig. \ref{fig-continuous} (c), we have also presented the wave functions in different regimes, showing that in the overlapped spectra, strong fluctuation of the wave functions is found. 
In contrast, in the extended phase, the wave functions are almost uniformly distributed in the space, and in the localized phase, they are localized exponentially in the small regime. 
All these features indicate criticality in the overlapped regime. 
It is necessary to point out that the continuous model has an obvious drawback that, to accurately calculate all the eigenvectors, only a small system can be dealt with.
Thus the large size limit of these critical states can not be reached in the current simulation. 
However, these results are sufficient to demonstrate the validity of the conclusion for CP using the general approach. 

\section{Conclusion and Discussion}
\label{sec-conclusion}

In this work, we present a general approach to realize the CP in a tunable way based on two coupled quasiperiodic chains. We demonstrate this phase using multifractal analysis. We show that the states in the overlapped spectra of localized states and extended states turn into critical and multifractal in the presence of inter-chain coupling.
By changing the forms of inter-chain coupling and quasiperiodic potentials, we demonstrate the existence of this CP to be rather general.
Especially, we also construct a
coupled chain with inter-chain self-duality. We demonstrate that the wave packet exhibits sub-diffusion dynamics when all states are critical and multifractal. Further, we also explain why the states in the overlapped regime are dramatically changed and the wave functions are reconstructed by the inter-chain coupling from the perspective of the divergence of perturbation series and effective unbounded potential. 
The physics of the states in the un-overlapped spectrum is also explained by the convergence of the perturbation series due to the finite gap. Finally, we construct a two-band  bichromatic incommensurate lattice using ultracold atoms and show that the CP may also be realized in the continuous models. 

Some remarkable applications of this mechanism to the CP are immediately feasible, which are outlined and briefly discussed below. 

(I) It can be used to realize the many-body CP by including many-body interaction, in which the states 
are expected to be extended yet non-thermal \citep{wang_realization_2020, cheng_many-body_2021}. In this literature, the authors found that the extended phase, localized phase, and CP in the single particle model can be driven to the many-body ergodic phase, many-body localized phase, and many-body CP, respectively, in the presence of many-body interaction.
In their models, the single-particle phase diagram and many-body phase diagram are found to be largely the same.
Since the single particle CP has been realized in our model in a controllable way, we expect it becomes the many-body CP in the presence of many-body interaction. 
The coupled models can provide another intriguing perspective for generating many-body CP. 
Let us assume that each chain has a conservation of the number of particles, then the Hilbert space of the two coupled chains follows $\mathcal{K}(N) = \oplus_{k=0}^N \mathcal{K}(k)$, with $\mathcal{K}(k) = \mathcal{K}_1(k) \otimes \mathcal{K}_2(N-k)$, where $\mathcal{K}_i(k)$ is the Hilbert space of $H_i$ with $k$ number particles. 
These subspaces are coupled consecutively by inter-chain coupling.
In this way, the overlap between the eigenstates in these subspaces can yield much more complicated
phases. We expect the overlapped spectra of $\mathcal{K}(k)$ and $\mathcal{K}(k\pm 1)$ can yield the CP. 

(II). Our model can greatly broaden the family of quasiperiodic models for the CP.
In the previous literature, as shown in the introduction, most of the models for the CP are based on Eq. \ref{eq-LforCP}, which can not be engineered and tuned easily in experiments.
In our approach, the relative energies of the two chains, and the inter-chain and intra-chain coupling 
can be tuned in experiments, thus it should be of great help for the realization and detection of the CP in experiments \cite{Xiao2021Observation}.
With this approach, one may also generalize this idea to many coupled chains, which may have much more interesting inter-chain dualities, thus with much more intriguing CPs. 

(III) Our approach may shed light on the possible underlying mechanism for the CP, which has not yet been fully understood. 
As discussed in the dilemmas in Sec. \ref{sec-mechansim}, in Sec. \ref{sec-approach}, and in Sec. \ref{sec-mechansim}, 
the interplay between localized states and extended states may turn the states into extended, localized, or critical, depending strongly on their competition.
For the random potential, it has been well established that any random potential can turn the extended states into localized states, we may only observe the localized states even in two coupled chains.
However, the quasiperiodic potential may lead to critical phases.
If this picture is right, it may provide a general approach for searching for CP on demand. 
More importantly, it may also be used to search for the CP in higher dimensions, in which extended states can still be survived with even random potential.
In this way, our understanding of the CP can be greatly enriched. 

To conclude, we expect much more investigation about this issue can be carried out in the future along this line for searching of more intriguing CPs.
We also anticipate the search for new physics in this fertile ground in the presence of many-body interaction and gauge phases. 

\begin{acknowledgements}
 We thank Tong Liu  and Yongping Zhang for the valuable discussion. This work is supported by the National Key Research and Development Program in China (Grants No. 2017YFA0304504 and No. 2017YFA0304103) and the National Natural Science Foundation of China (NSFC) with No. 11774328.
\end{acknowledgements}

\appendix
\section{Critical states in the spin-orbital couple lattice system}
\label{sec-appendix-soc} 
Recently, Wang \textit{et al.} \citep{Wang2022Quantum} found 
a quantum phase with three coexisted energy-dependent regimes, {\it i.e.}, the extended, critical, and localized phases, which can be realized in a spin-orbit coupled lattice model with quasiperiodic potential. It
was claimed that spin-orbital coupling can make the critical point into a critical regime for the CP. Here, we emphasize that this critical regime corresponds to the CP discussed in this manuscript. 
In this appendix, we will briefly discuss the CP in their model, showing that this finding can also be incorporated into our picture for CP. The spin-orbital coupled Hamiltonian reads as 
\begin{equation}
\begin{aligned}
H_0 & =-t_0 \sum_{\langle i, j\rangle}\left(c_{i, \uparrow}^{\dagger} c_{j, \uparrow}-c_{i, \downarrow}^{\dagger} c_{j, \downarrow}\right) \\
& + t_\text{so} \left[\sum_i\left(c_{i, \uparrow}^{\dagger} c_{i+1, \downarrow}-c_{i, \uparrow}^{\dagger} c_{i-1, \downarrow}\right)+\mathrm{h.c.}\right] \\
& +\sum_i \delta_i\left(n_{i, \uparrow}-n_{i, \downarrow}\right)+\lambda \sum_i \delta_i\left(n_{i, \uparrow}+n_{i, \downarrow}\right),
\end{aligned}
\label{eq-appendix-soc}
\end{equation}
with $\delta_j = V \cos(2\pi\alpha j)$, $V$ is the strength of inter-chain coupling, and $t_\text{so}$ is the spin-orbit coupling strength, which plays the same role as $H_c$. We set $t_0 = 1$ as the energy scale.
This model can be viewed as two coupled chains if we rewrite it as 
\begin{equation}
\begin{aligned}
H_{0, \uparrow} &=\sum_i\left[-t_0\left(c_{i, \uparrow}^{\dagger} c_{i+1, \uparrow}+\text{h.c.} .\right)+(1+\lambda) \delta_i n_{i, \uparrow}\right], \\
H_{0, \downarrow} & = - \sum_i\left[-t_0\left(c_{i, \downarrow}^{\dagger} c_{i+1, \downarrow}+ \text{h.c.}\right)+(1-\lambda) \delta_i n_{i, \downarrow}\right], \\
H_\text{so} &= t_\text{so} \left[\sum_i\left(c_{i, \uparrow}^{\dagger} c_{i+1, \downarrow}-c_{i, \uparrow}^{\dagger} c_{i-1, \downarrow}\right)+ \mathrm{h.c} .\right].
\end{aligned}
\end{equation}
Then the two decoupled models
$H_{0\uparrow}$ and $H_{0\downarrow}$ are the AAH model with different quasiperiodic potential strengths, and $H_\text{so}$ plays the same role as $H_c$ for inter-spin coupling. 

\begin{figure}[!htbp]
\includegraphics[width=0.48\textwidth]{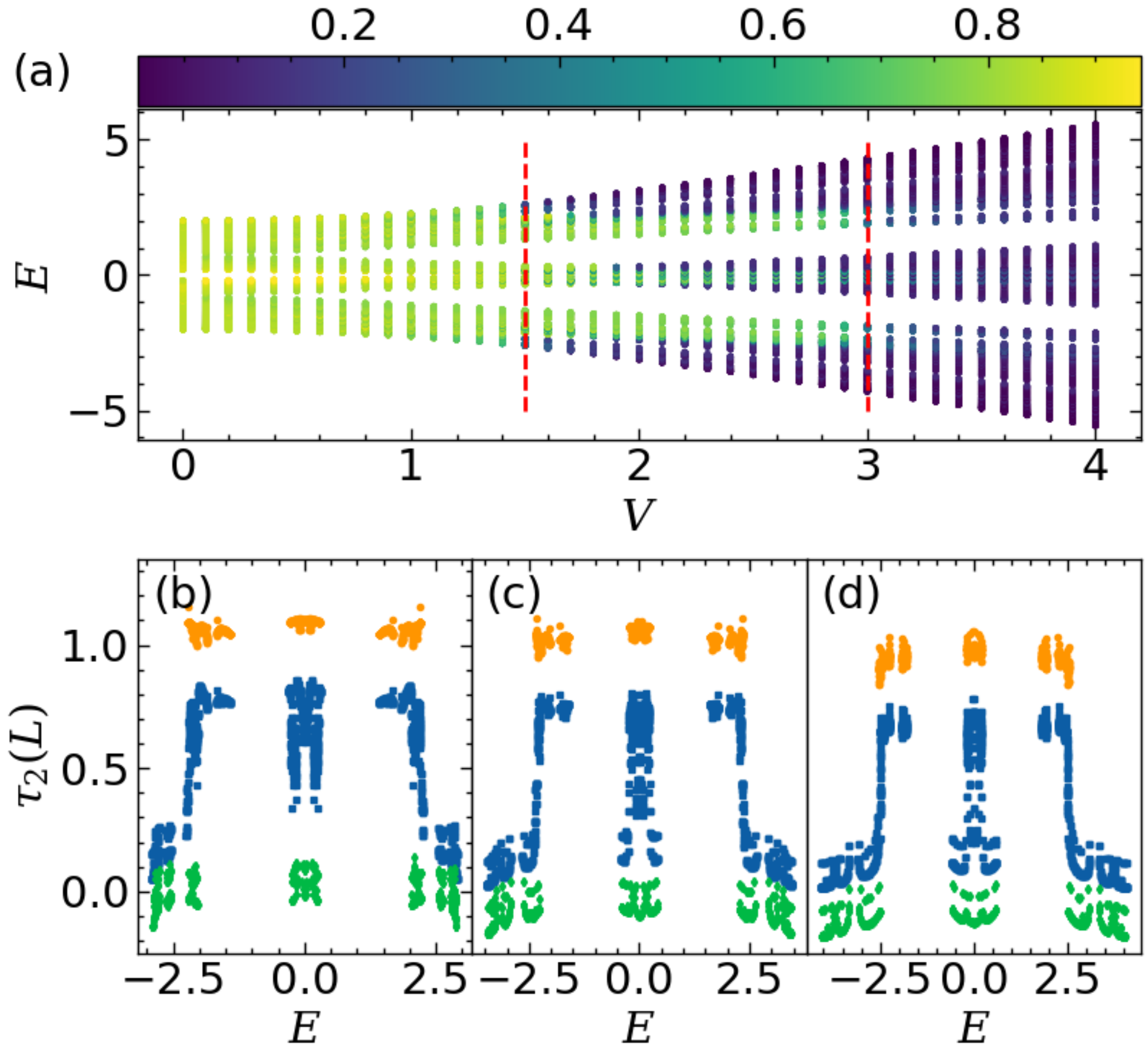}
\caption{
(a) The FD $\tau_2(L)$ for the states in the spin-orbital coupled model in Eq. \ref{eq-appendix-soc} with $L= 2F_{15} = 1220$, $t_{\mathrm{so}}=0.1$, $\lambda=1/3$ for different incommensurate lattice strength $V$; notice that $t_0 =1$. The red dashed lines denote $V_{c1} = 2/(1+\lambda)$ and $V_{c2} = 2/(1-\lambda)$, which are the boundaries for the critical states.
(b)-(d) The blue squares are FD $\tau_2(L)$ for the states in Eq. \ref{eq-appendix-soc} with (a) $V = 1.8$; (b) $V = 2.3$; (d) $V = 2.8$. The yellow dots are the FD of states of $H_{0,\downarrow}$ with disorder strength $(1-\lambda)V$ while the green diamonds are the FD of states of $H_{0,\uparrow}$ with disorder strength $(1+\lambda)V$.
The CP emerges from the overlapped spectra if we switch on the inter-chain couplings (see the blue squares).
}
\label{fig-soc}
\end{figure}

As has been discussed in the main text, we calculate the FD $\tau_2(L)$ as a function of $t_\text{so}$, which
are presented in Fig. \ref{fig-soc} (a), showing that in the weak coupling regime, the overlapped spectra between
these two components will become critical. In the large $t_\text{so}$, the repulsive interaction between the two 
components leads to localized and extended states. We find that the overlapped regime can be realized when 
$2/(1+\lambda)<V<2/(1-\lambda)$, which is predicted in Fig. \ref{fig-soc} (a). A cross-section of this diagram for different $V$ is presented in Fig. \ref{fig-soc} (b) to (d), all of which show the CP in the overlapped regime, while in the un-overlapped regime, $\tau_2$ equals either unity or zero.
Therefore, we conclude that our theory can also be applied to study the CP phase in the above spin-orbit coupled models, which also manifests the generality of our approach for CP. 

\section{Possible scheme for experimental realization in spin-dependent bichromatic incommensurate lattice}
\label{sec-appendix-experiment}

The coupled chain model has been realized in many experimental systems, including cold atoms in spin-dependent potential 
\citep{mandel_coherent_2003,yang_spin-dependent_2017}, optical lattice with two-leg ladders \citep{atala_observation_2014, Atala2014Meissner, Livi2016Synthetic} {\it etc.}.
The spin-orbit coupled model above can also be regarded as one possible candidate for CP. 
Here, we proposed a scheme to realize the coupled quasiperiodic chains using a two-component ultracold bosonic $^{\mathrm{87}}\mathrm{Rb}$ atoms in a spin-dependent incommensurate optical lattice, where the potential in each chain can be tuned individually.
We choose the hyperfine states $|0\rangle = |F = 1,m_F = -1\rangle$ and $|1\rangle = |F = 2,m_F = -2\rangle$ to represent the two chains.
The primary optical lattice $V_\text{p}(z)$, the second incommensurate optical lattice $V_{\text{in},j}(z)$, and inter-chain coupling are realized using the following ways.

\begin{figure}[!htbp]
\centering
\includegraphics[width=0.45\textwidth]{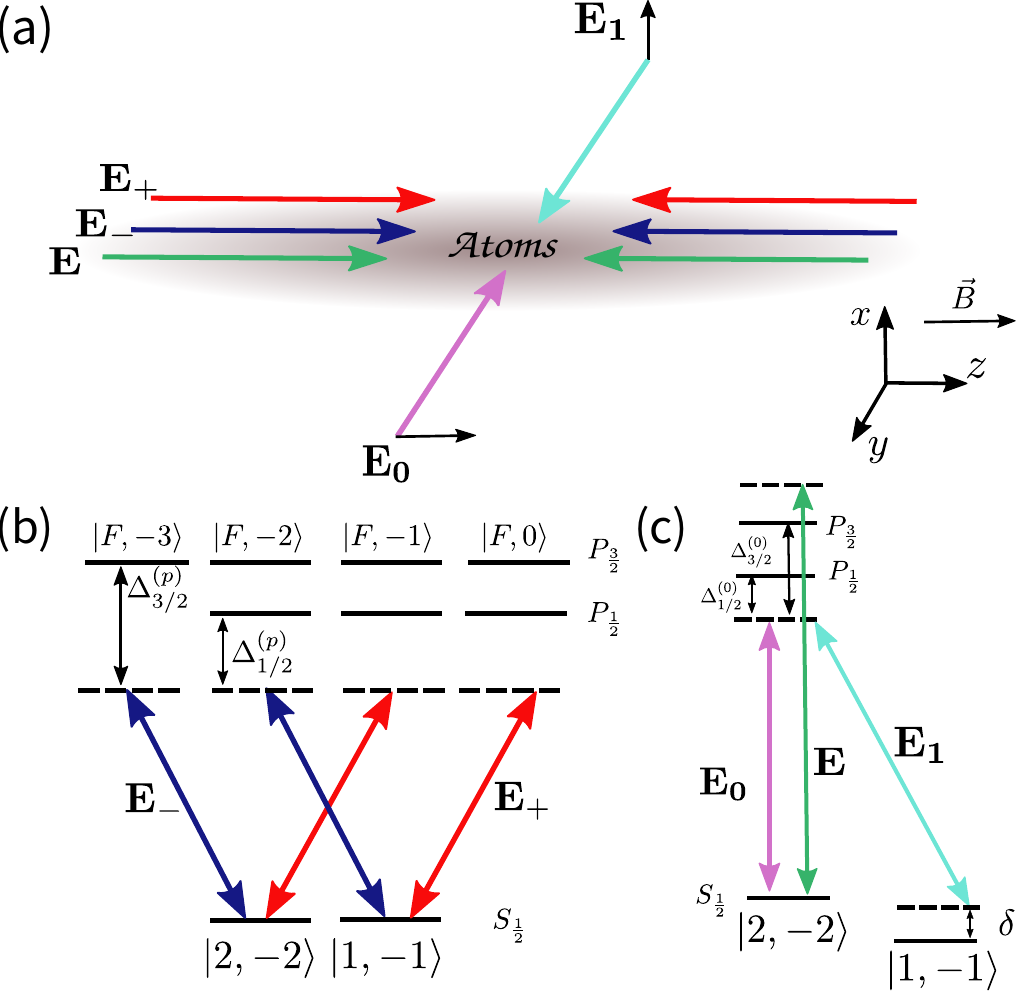}
\caption{
Experimental setup and light couplings for $^{87}\mathrm{Rb}$ atoms. 
(a). Schematic of the experimental setup. A pair of counter-propagating coherent waves with left-handed polarization forms a standing wave $\mathbf{E_+}$ of frequency $\omega'$.
The same approach is used to generate standing wave $\mathbf{E_-}$ with right-handed polarization of frequency $\omega'$.
The counter-propagating waves $\mathbf{E}$ with linear polarization form the spin-independent primary lattice with wave-vector $k=k_1$.  
Then, we use a plane wave $\mathbf{E_0}$ along the y-axis with $z$ polarization and a plane wave $\mathbf{E_{1}} $ with $x$ polarization to induce a Raman process, which plays the roles of inter-chain coupling.
(b). Optical transitions for generating the incommensurate lattice are the same as that of the incommensurate lattice.
(c). Optical transition for Raman process and for generating primary lattice.
}
\label{fig-experiment-scheme}
\end{figure}

1. Realization of the spin-independent primary optical lattice. Here, $V_\text{p}(z)$ is generated by a standing polarized wave field $E = \frac{E}{2}\mathbf{e_x}\cos(k_\text{p} z - \phi_\text{p})$ of frequency $\omega_\text{p}$, which is schematically shown in Fig. \ref{fig-experiment-scheme} (c).
This standing wave generates a scalar energy shift with lattice constant $a = \pi/k_\text{p}$, which reads as 
\begin{equation}
V_\text{p}(z) \propto \cos^2(k_\text{p} z - \phi_\text{p}).
\end{equation}
This scheme for generating spin-independent optical lattice has been applied in various experiments \citep{Greiner2002Quantum, Fallani2004Observation, Morsch2006Dynamics}, thus are not discussed in detail.  

2. Realization of the spin-dependent incommensurate optical lattice. The incommensurate potential is generated by a standing polarized wave field $\mathbf{E_+} = \frac{E_+}{2}\mathbf{e_{+}}\mathrm{e}^{\mathrm{i}(\phi_+ + \phi_{+e})}\cos(k_\text{in}z - \phi_{1e})$ and a standing polarized wave $\mathbf{E_-} = \frac{E_-}{2}\mathbf{e_{-}}\mathrm{e}^{\mathrm{i}(\phi_- + \phi_{-e}}\cos(k_\text{in}z - \phi_{-e})$, with the same frequency $\omega$ (see Fig. \ref{fig-experiment-scheme} (b)), where $\phi_{+,-}$ and $\phi_{+e}$, $\phi_{-e}$ are phases that can be controlled in experiment. 
We note that $ \mathbf{e_{+}} = \mathbf{x} + i\mathbf{y}$ and $\mathbf{e_{-}} = \mathbf{x} - i\mathbf{y}$, which can be realized by two coherent counter-propagating waves \citep{fang_coherent_2016} as shown in Fig. \ref{fig-experiment-scheme} (a). These two standing waves couple different hyperfine states and generate a spin-dependent scalar potential to the state $|0\rangle$ or $|1\rangle$. The scalar potential is  
\begin{eqnarray}
V_{\text{in},j}(z) && = E_{+}^2 A_{+,j} \cos^2(k_\text{in} z - \phi_{+}) \nonumber \\ && +  E_{-}^2 A_{-,j} \cos^2(k_\text{in} z - \phi_{-}),
\label{eq-incommensurate-p}
\end{eqnarray} 
with $ A_{+,1} = \frac{|t_{3/2}|^2}{\Delta_{3/2}^{(p)}}\frac{1}{6} + \frac{|t_{1/2}|^2}{\Delta_{1/2}^{(p)}}\frac{2}{3} $, $ A_{-,1} = \frac{|t_{3/2}|^2}{\Delta_{3/2}^{(p)}}\frac{1}{2} $, $ A_{+,0} = \frac{|t_{3/2}|^2}{\Delta_{3/2}^{(p)}}\frac{5}{12} + \frac{|t_{1/2}|^2}{\Delta_{1/2}^{(p)}}\frac{1}{6} $ and $ A_{-,0} = \frac{|t_{3/2}|^2}{\Delta_{3/2}^{(p)}}\frac{1}{4} + \frac{|t_{1/2}|^2}{\Delta_{1/2}^{(p)}}\frac{1}{2} $. 
In these coefficients, we have defined the dipole matrix elements $t_{1 / 2} = \left\langle J=1 / 2\|e \mathbf{r}\| J^{\prime}=1 / 2\right\rangle$ and $t_{3 / 2} =\left\langle J=1 / 2\|e \mathbf{r}\| J^{\prime}=3 / 2\right\rangle$, which approximate to $t_{3/2} \simeq \sqrt{2}t_{1/2}$.
Then we have 
\begin{equation}
\begin{aligned}
A_{+,1} & = |t_{1/2}|^2\left(\frac{1}{3\Delta_{3/2}^{(p)}} + \frac{2}{3\Delta_{1/2}^{(p)}}\right), \\
A_{-,1} & = \frac{|t_{1/2}|^2}{\Delta_{3/2}^{(p)}},
\end{aligned}
\label{eq-cof-a1}
\end{equation}
and
\begin{equation}
\begin{aligned}
       A_{+,0} &= |t_{1/2}|^2\left(\frac{5}{6\Delta_{3/2}^{(p)}} + \frac{1}{6\Delta_{1/2}^{(p)}}\right), \\
       A_{-,0} &= |t_{1/2}|^2 \left( \frac{1}{2\Delta_{3/2}^{(p)}} + \frac{1}{2\Delta_{1/2}^{(p)}} \right).
\end{aligned}
\label{eq-cof-a2}
\end{equation}
Since $A_{\pm,j}$ are different for index $j=0$ and $j = 1$, the scalar potential in Eq. \ref{eq-incommensurate-p} can be controlled individually for each state $|j\rangle$, which is crucial for realizing critical states.
For $^{87}$Rb atoms, the energy between difference manifold $P_{\frac{3}{2}}$ and $P_{\frac{1}{2}}$ is about 29.4584 meV. We may choose the wavelength of the wave fields $\mathbf{E}_+$ and $\mathbf{E}_-$ as 804 nm, which is red detuned (see Fig. \ref{fig-experiment-scheme} (b)).
Then the value pf the detunings are $\Delta_{1/2}^{(p)} \sim 19.4$ meV and $\Delta_{3/2}^{(p)} \sim 48.9$ meV. 
We neglect the constant terms in Eq. \ref{eq-incommensurate-p} and
set $\phi_+ = 0$ and $\phi_- = \pi/4$ for simplification.
The incommensurate potential in Eq. \ref{eq-incommensurate-p} becomes
\begin{equation}
    V_{\text{in},j} = \frac{B_{j}}{2} \cos(2k_\text{in} + \theta_{j}),
\end{equation}
with $B_j = \sqrt{E_+^4A_{+,j}^2 + E_-^4 A_{-,j}^2}$ and $\theta_j =\arctan( E_+^2A_{+,j}/E_-^2A_{-,j} )$.
It follows that the ratio between these two potential depths can be written as
\begin{equation}
    \frac{|B_0|}{|B_1|} \simeq \sqrt{\frac{A_{+, 0}^2 (E_+/E_-)^4+ A_{-, 0}^2}{A_{+, 1}^2 (E_+/E_-)^4 + A_{-, 1}^2}}.
\end{equation}
Combining Eq. \ref{eq-cof-a1}, Eq. \ref{eq-cof-a2}, and the value of the detuning $\Delta_{1/2}^{(p)}$ and $\Delta_{3/2}^{(p)}$, we will find that 
\begin{equation}
0.63<\frac{|B_0|}{|B_1|}<1.761,
\end{equation}
which is hopeful for realizing a wide parameter area for the CP.

\begin{figure}[!htbp]
\centering    
\includegraphics[width=0.48\textwidth]{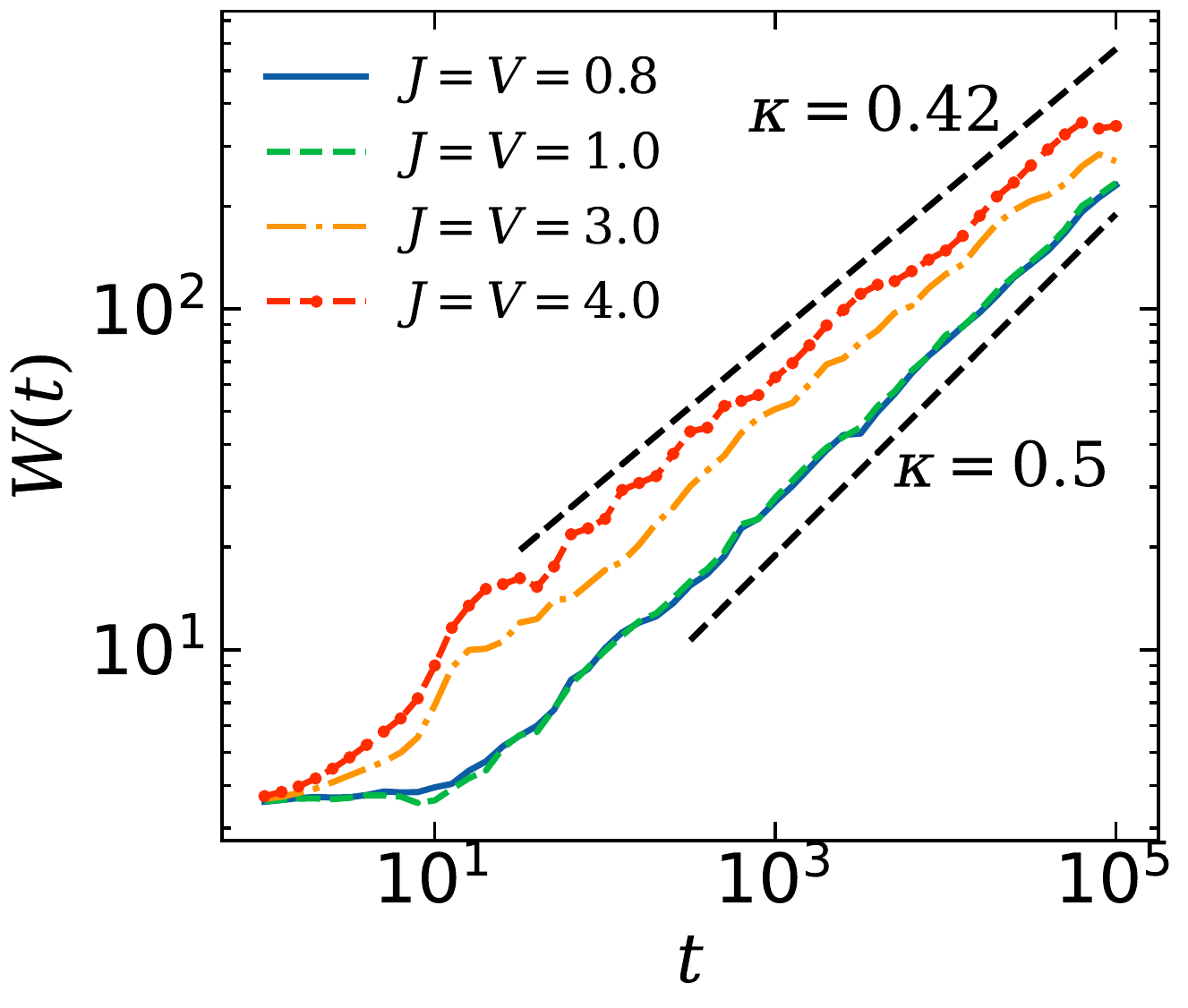}
\caption{The width $W(t)$ versus time $t$ for different $J$ and $V$. We use $L = 2F_{17} = 3194$ and $t_\text{v}=0.5$. The initial states in chosen to be $\gamma = \text{loc}$ with its form the same as Eq. \ref{eq-wave-packet-initial}. The black dashed lines denote the function $ 0.6 t^{0.5}$ and $(6/\sqrt{2})t^{0.42}$. }
    \label{fig-wave-packet}
\end{figure}

3. Realization of the inter-chain coupling by Raman coupling. We use plane waves $\mathrm{E_0} = \mathbf{e_z}\exp\{-\mathrm{i}k_0 y - \mathrm{i}\omega_0 t\}$ and $\mathrm{E_1} = \mathbf{e_x}\exp\{\mathrm{i}k_1 y - \mathrm{i}\omega_1 t\}$ to generate a two-photon Raman process for the
inter-chain coupling (see Fig. \ref{fig-experiment-scheme} (c)). Using the two plane waves, the scalar shift of the atoms, which is a uniform constant, can be eliminated. In addition, with the two-photon detuning defined as $\delta = \omega_1 - \omega_0$, we can turn the energy offset of the two chains to control the overlapped spectra. The coupling strength between the hyperfine states is given by  
\begin{equation}
t_\text{v} \sim E_0 E_1\left(\frac{1}{\sqrt{12}}\frac{|t_{1/2}|^2}{\Delta_{1/2}^{(0)}} + \frac{1}{\sqrt{24}}\frac{|t_{3/2}|^2}{\Delta_{3/2}^{(0)}} \right) \mathcal{A}, 
\end{equation}
where $\mathcal{A}$ is the overlap of the functions between the two chains. Compared with the coefficients of $A_{\eta,j}$ and $t_\text{v}$, we see that these parameters can be tuned independently to some extent. It is also necessary to emphasize that the coupling between the two hyperfine states may also be realized using microwaves \citep{Gupta2003Radio, Meng2023Atomic}. 

\section{Wave packet dynamics of the coupled dual quasiperiodic chains}
\label{sec-appendix-wavepacket}

In this appendix, we discuss the dynamics of wave packet diffusion in coupled dual quasiperiodic chains with other dual parameters. The results for various $J=V$ are presented in Fig. \ref{fig-wave-packet}. We find that the dynamical exponent $\kappa \sim 0.42$ for the system with $J = V = 3$ and $J = V = 4$; $\kappa \sim 0.5$ for the system with $J = V = 0.8$ and $J = V = 1.0$. Thus this exponent may depend weakly on the system parameters, all of which are around the value of $\kappa=1/2$. A similar conclusion was reported in the Frenkel-Kontorova model \citep{Tong2002Electronic}, where the dynamical exponent also depends weakly on the system parameters.

\bibliography{ref}

\end{document}